\PassOptionsToPackage{pdfpagelabels=false}{hyperref}

\documentclass[preprint,authoryear,12pt]{elsarticle}  

\usepackage{graphicx}
\usepackage{amsmath}
\usepackage{epsfig}
\usepackage[latin1]{inputenc}
\usepackage{natbib}
\usepackage{amsmath} 
\usepackage{amssymb} 
\usepackage{lineno}

\usepackage[T1]{fontenc}
\usepackage{ae,aecompl,url,wasysym}
\usepackage{multicol}
\newcommand\NS{N\!S}

\title{Near/Far Side Asymmetry in the Tidally Heated Moon}

\author[a1]{Alice C. Quillen\corref{cor1}}
\ead{alice.quillen@rochester.edu}
\author[a1,a2]{Larkin Martini}
\ead{lmartini@mymail.mines.edu}
\author[a1,a3]{Miki Nakajima}
\ead{mnakajima@rochester.edu}

\address[a1]{Department of Physics and Astronomy, University of Rochester, Rochester, NY 14627, USA}
\address[a2]{Department of Geology and Geological Engineering, Colorado School of Mines, Golden, CO 80401, USA}
\address[a3]{Department of Earth and Environmental Sciences, University of Rochester, Rochester, NY 14627, USA}
\cortext[cor1]{Corresponding author}

\begin{document}

\begin{abstract}

Using viscoelastic mass spring model simulations to  track heat distribution inside a tidally perturbed body, we measure the near/far side asymmetry of heating in the crust of a spin synchronous Moon in eccentric orbit about the Earth. With the young Moon within  $\lesssim 8$ Earth radii of the Earth, we find that tidal heating per unit area in a lunar crustal shell  is asymmetric due to the octupole order moment in the Earth's tidal field and is 10 to 20\% higher on its near side  than on its far side. Tidal heating reduces the crustal basal heat flux and  the rate of magma ocean crystallization. Assuming that the local crustal growth rate depends on the local basal heat flux and the distribution of tidal heating in latitude and longitude, a heat conductivity model illustrates that a moderately asymmetric and growing lunar crust could maintain its near/far side thickness asymmetry but only while 
the Moon is near the Earth. 



\end{abstract}


\maketitle

\section{Introduction}

Gravity anomaly maps show that the Moon exhibits a crustal dichotomy 
with a crust that is about twice as thick (50--60 km)
on the Earth's far side as on the nearside (20--30 km) 
\citep{zuber94,wieczorek13}.
Lunar crustal rocks are extremely old, many with
ages within the first 200 Myr after the formation of the first solids in the solar system; 
for reviews see \citet{borg99,mccallum01,shearer06,elkinstanton12}. 
Evidence that there was a lunar magma ocean, and that it solidified fractionally
rather than in bulk, comes from measurements of crustal composition, 
ages of crustal rocks and trace elements in suites of rocks (including KREEP;
the potassium, rare-earth elements and  phosphorus rich geochemical component of some lunar rocks).
As the lunar magma ocean solidified, the remaining liquid became 
progressively enriched in incompatible elements. 
The lunar magma ocean solidified to 80\% in  approximately a thousand years 
after the formation of the Moon  \citep{elkinstanton11}, at which 
time plagioclase began to crystallize.
The plagioclase crystals were lower density than the surrounding magma 
and so rose to form the anorthitic crust \citep{wood70,snyder92,abe97}.
When plagioclase crystallization began,  the remaining lunar magma ocean would have been about  100  km deep
\citep{elkinstanton11}.
Heat was probably transferred conductively through the lunar crust,
and convectively in the lunar magma ocean.  
Once a solid crustal lid was formed,  magma ocean cooling slowed, taking
 about 200 Myr  to completely solidify \citep{elkinstanton11}, 
with tidal heating serving as an additional crustal heat source that significantly delayed (by a factor of about 10) 
the time for the lunar magma ocean to completely solidify  \citep{meyer10,elkinstanton11}.

A number of models have been proposed to account for the Moon's crustal dichotomy.
For example, \citet{jutzi11}  propose that the asymmetry 
formed by accretion of two moonlets that were formed by a larger previous impact.
However,  the high magnesium fraction compared to iron  in the far side lunar crust
compared to the nearside crust  suggests that the crustal dichotomy 
occurred during its formation, and with the far side forming earlier than the nearside \citep{ohtake12}.
Asymmetric heating from earthshine could have induced
a tilted global convection pattern in the magma ocean, where parcels rise and sink at an angle from the vertical,
and this could have caused uneven crust growth due to crystal transport from the near
to farside \citep{loper02}. 
Chemical stratification could cause large wavelength gravitational instability
that causes a thick dense crustal layer to grow  \citep{parmentier02}.  
\citet{wasson80} explored the role of Earthshine (also see \citealt{roy14}), 
asymmetric thermal insulation by the crust
(a floating continent), asymmetric bombardment and an asymmetric internal core,  
discarding all but the asymmetric core model.  
In their floating continent crustal insulation model, asymmetric crystallization
 in the lunar magma ocean is caused by differences in the 
local cooling rate associated with a detached
and floating anorthosite continent that insulates the magma ocean beneath it. 
 Inefficient lateral mixing in the magma ocean is required to allow temperatures 
beneath the floating continent to be appreciably higher than those in the opposite hemisphere.   
Lunar magma ocean dynamical models estimate a high Rayleigh number 
(up to $Ra \sim 10^{25}$) for the early lunar magma ocean \citep{spera92}, implying
that it is well mixed.  Mixing caused by vigorous convection presents
a challenge for models of asymmetric lunar crustal growth.

\subsection{Event timeline}
\label{sec:timeline}

\begin{figure*}
\centering
    \includegraphics[width=5in]{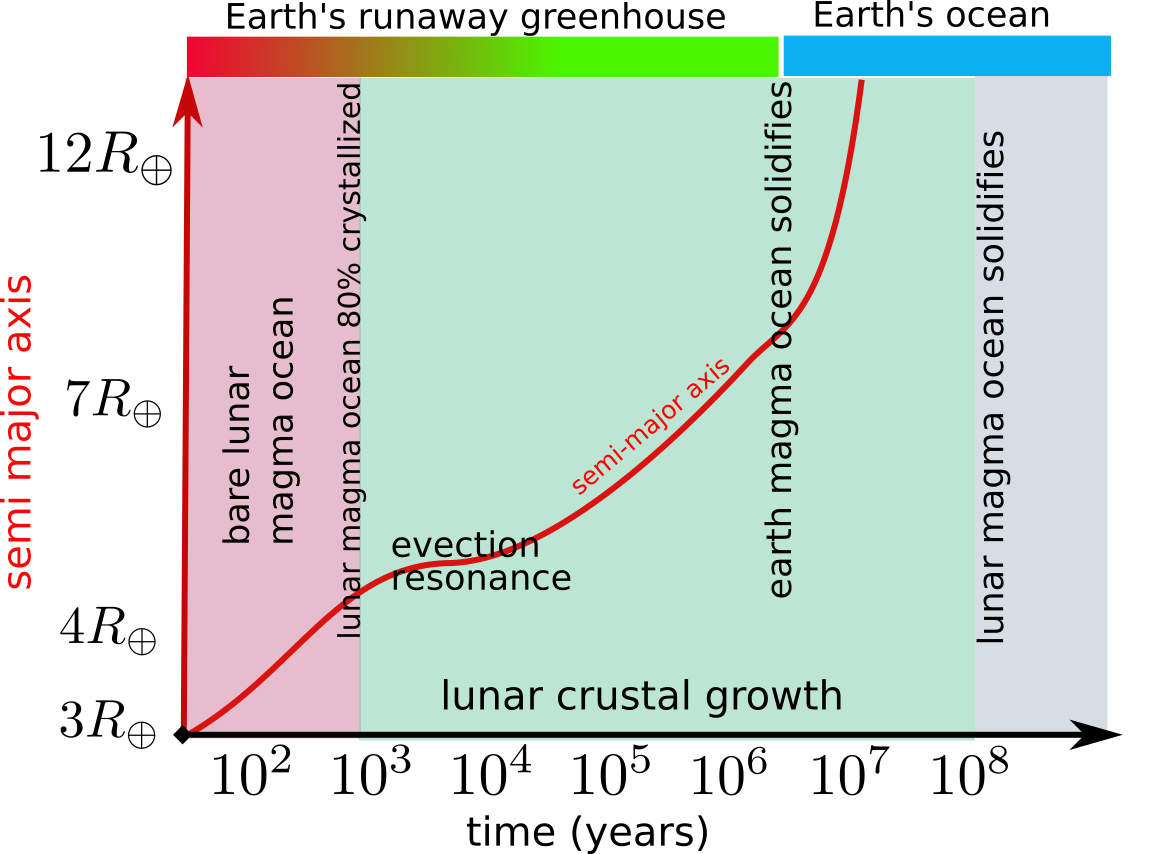} 
\caption{
A  timeline for processes occurring after the formation of the Moon.  
The $x$ axis shows time since Moon formation.
The red line approximately shows the Moon's semi-major as a function of time in units
of Earth radii using the left-hand side's $y$ axis.
The color bar on the top shows the Earth's effective radiative temperature.
Background colors show the state of the Moon, with pink denoting a bare magma lunar surface,
green the presence of a crust overlaying a magma ocean and blue after solidification.
The Earth's magma ocean solidified
after few Myr (at the latest 6 Myr) \citep{zahnle07,zahnle15}.  
We show slow tidal drift in the Moon's semi-major axis assuming that the tidal dissipation $Q$ for Earth
is large when the Earth hosted a magma ocean, prior to a few million years after Moon formation \citep{zahnle15}. 
After the Earth's magma ocean froze, tidal dissipation in the Earth increases, increasing
the drift rate of the Moon's semi-major axis.
Prior to solidification of the Earth's magma ocean, the Earth hosted a runaway greenhouse
atmosphere with radiative temperature slowly declining, eventually reaching the effective temperature set
by radiation balance with the Sun; $T_\oplus \sim 280^\circ$ K. 
A level of 80\% crystallization in the lunar magma ocean
is reached in $\sim 10^3$ years after Moon formation \citep{elkinstanton11} at which 
time a buoyant  plagioclase crust begins to grow, slowing the cooling rate of the Moon's magma ocean.
Passage through evection resonance at a semi-major axis of about  $5 R_\oplus$
can increase the Moon's orbital eccentricity but
may also decrease the Moon's semi-major axis \citep{touma98,tian17}.
The orbital eccentricity could also increase afterwards \citep{zahnle15}.
For alternate scenarios see \citet{zahnle15,tian17}.
Tidal heating on the Moon is strongly dependent on its orbital eccentricity and semi-major axis  and decreases
as its semi-major axis increases.
 \label{fig:timeline}}
 \end{figure*}

We  summarize a rough timeline for events
 occurring during the formation and growth
of the Moon's crust, according to prevailing current understanding (see Figure \ref{fig:timeline}).
For more details on this timeline and alternate scenarios,  see \citet{cuk12,zahnle15,cuk16,tian17}.
The ages of the oldest terrestrial and lunar zircons imply that 
 the Earth's surface must have cooled before
crystallization of the lunar magma ocean was complete \citep{nemchin09,elkinstanton12}. 
Because of its water content, the Earth would have formed a thick greenhouse atmosphere
a few thousand years after Moon formation \citep{zahnle07,sleep14,lupu14,zahnle15}.  The optically thick
atmosphere slows the cooling of the Earth and reduces its effective radiative temperature.
The Moon is expected to have spun down into  a spin synchronous or tidally locked 
state 1--100 years after formation \citep{garrickbethell06,garrickbethell14}.

For a few thousand years,  the Earth's radiative temperature would be  a few hundred degrees higher
than the effective temperature set by the Solar constant (see Figure 3 by \citealt{zahnle15}).
During this time Earthshine could cause moderate  heating on the Earth's near side compared
to the far side \citep{roy14}.   
Equations 2 and 3 by \citet{roy14}  estimate the temperatures
of the near and far sides of the Moon
\begin{align}
T_{\rm far} &= \left[ \frac{T_\odot^4 R_\odot^2}{4 D_\odot^2} \right]^\frac{1}{4} \\
T_{\rm near} &= \left[ \frac{T_\odot^4 R_\odot^2}{4 D_\odot^2}  + \frac{T_\oplus^4 R_\oplus^2}{2 D_\oplus^2} \right]^\frac{1}{4}
\end{align}
where $D_\odot, R_\odot, T_\odot$ are the distance to the Sun, and the radius and effective temperature of the Sun,
$D_\oplus, R_\oplus, T_\oplus$ are the distance of the Moon from the Earth, and the  
radius and effective temperature of the Earth.
We estimate that 
at distance of $D_\oplus = 4 R_\oplus$ and with the Earth's radiative temperature at $500^\circ$ K, 
the near side of the Moon would be about $11^\circ$ K hotter than the far side.
This is 4\% of the Moon's effective surface temperature of about $280^\circ$ K, set by the Solar constant.

The Earth's greenhouse atmosphere
lasts until the Earth's magma ocean freezes, or a few million years after Moon formation
 \citep{zahnle07,sleep14,lupu14,zahnle15}.  
At this time the tidal dissipation rate in the Earth drops from $Q_\oplus \gtrsim 10^4$ to $Q_\oplus \lesssim 100$ \citep{zahnle15}. 
A higher $Q_\oplus$ implies a slower drift rate of the Moon's semi-major axis \citep{sleep14,zahnle15}.
This scenario is called the `tethered Moon' scenario.
With such a high value of $Q_\oplus$, coupled tidal evolution and 
 thermal modeling imply that the Moon must solidify prior to reaching a semi-major axis of $20 R_\oplus$ in order
to maintain its orbital inclination going through the Cassini state transition at $30 R_\oplus$  \citep{chen16}.

\citet{touma98} showed that a resonance between perigee and perihelion, the {\it evection} resonance,
could increase the Moon's orbital eccentricity. 
The evection resonance occurs when the precession period of the Moon's pericenter 
(due to torque from the oblate Earth) is equal to the Earth's orbital period.
The evection resonance location given by $a_{\leftmoon}$, the semi-major axis of the Moon's orbit
about the Earth, depends on the semi-major axis of the Earth's orbit
around the Sun, $a_\oplus$,  and the Earth's  gravitational moment, $J_{2\oplus}$.
The resonance occurs where
\begin{equation}
n_\oplus \approx \frac{3}{2} J_{2\oplus} \frac{R_\oplus^2}{a_{\leftmoon}^2} n_{\leftmoon},
\end{equation}
where $n_\oplus$ is the Earth's mean motion in orbit about the Sun (center of mass of
Earth/Moon binary), $n_{\leftmoon}$ is the Moon's mean motion about the Earth,  
and the mean radius of the Earth is $R_\oplus$.
\citet{touma98} estimated the location of the eviction resonance assuming that the Earth's gravitational moment
 $J_{2\oplus} \propto \Omega_\oplus^2$, proportional to the square of the 
Earth's rotation rate.  This assumption is consistent with a core and mantle hydrostatic model 
\citep{dermott79b}.
The resonance condition 
\begin{equation}
\frac{a_{\leftmoon {\rm evec}}}{R_\oplus} \approx \left( \frac{M_\oplus}{M_\odot} \right)^\frac{1}{7} 
 \left( \frac{3}{2} J_{2\oplus} \right)^\frac{2}{7} \left(\frac{a_\oplus}{R_\oplus} \right)^\frac{3}{7}
\end{equation}
where $a_{\leftmoon, {\rm evec}}$ is the orbital semi-major axis of resonance.
With the Earth's rotation period of 5.2 hours, scaling from the Earth's current $J _{2\oplus} \sim 10^{-3}$
gives a location for the resonance
$a_{\leftmoon {\rm evec}} \sim 4.6 R_\oplus$, as estimated by \citet{touma98}.
However, if the Earth's spin period is only 3 hours, as in a fast spinning Earth scenario \citep{cuk12},
the evection resonance is more distant, at $a_{\leftmoon {\rm evec}} \sim 6.4 R_\oplus$.

If the Moon's semi-major axis is drifting slowly, due to a high $Q_\oplus$, capture into the evection
resonance is assured, however the ratio of tidal heating parameters (known as the Mignard $A$ parameter) would limit the orbital eccentricity growth while in the evection resonance \citep{zahnle15}.  A non-zero but low eccentricity
might have been present when the Moon was near the Earth.  The non-zero eccentricity allows
the Moon to be tidally heated during this time.
\citet{zahnle15} find that the eccentricity can grow to higher values  when $a_{\leftmoon}$ is between 20 and 40 $R_\oplus$
 due to the higher tidal dissipation in the Earth after it has solidified.   

The Moon cannot coalesce within a semi-major axis of $2.9 R_\oplus$ \citep{kokubo00}
equal to the Roche limit. The Roche limit is fairly large 
due to the higher mean density of Earth compared to that of the Moon.  
Because the evection resonance is located at a semi-major axis of only a few Earth radii,
passage through  evection resonance  took place soon after Moon formation. 
Passage through evection resonance 
may also decrease the Moon's semi-major axis \citep{tian17,touma98}.
The evection resonance
may have reduced the total angular momentum of the Earth-Moon system \citep{cuk12}.
The Moon can escape the evection resonance if the tidal drift rate is fast enough to be non-adiabatic 
\citep{touma98} or if tidal parameters (Love number and dissipation factor)
vary and in this case the Moon could have
passed through the resonance multiple times \citep{tian17}.   \citet{zahnle15} estimate
that the evection resonance is encountered sufficiently early that the Earth would be molten during the encounter.
The Earth's molten state implies that 
tidal dissipation in the Earth would be lower than estimated for a solid Earth
giving slower orbital evolution  \citep{zahnle15}.  Thus tidal evolution 
could have taken place quite slowly in the vicinity of the evection resonance, and when the eccentricity
was non-zero.

In summary, a magma ocean on the Earth may have reduced the drift rate in semi-major axis of the Moon, prolonging
its passage through the evection resonance and allowing the lunar magma ocean to solidify when
the Moon's orbit was eccentric enough to be tidally heated and when the Moon was quite near the Earth.

\subsection{Tidal heating when the Moon is near the Earth}
\label{sec:tide}

In this paper we reconsider the notion that the lunar crust grew asymmetrically, rather
than acquired its dichotomy after formation (as in the impact scenario by \citealt{jutzi11}).
In icy satellites with a crust decoupled from the mantle by a subsurface ocean, 
tidal dissipation in the crust is enhanced and can lead to global crustal thickness variations due to 
differences in the spatial distribution of tidal heating \citep{ojakangas89,tobie03,tobie05,nimmo07,garrickbethell10}. 
In the absence of strong tidal heating, the equilibrium crustal thickness would be set by 
 the equilibrium surface temperature, which varies as a function of latitude.  This is ruled out
 by crustal thickness measurements, implying that additional processes, such
 as tidal heating must have affected crustal growth \citep{garrickbethell10}.
\cite{garrickbethell10} proposed that degree 2 structure in the lunar highlands topography
could be due to tidal heating that was enhanced near the poles.

\cite{garrickbethell10}  computed the tidal heating rate using the quadrupole term in  the expansion of the 
gravitational potential from the tidal perturber (here the Earth), as is commonly done because
 this term is the strongest \citep{kaula64,peale78}.  
The gravitational potential  of
a point mass perturber (the Earth) of mass $M_\oplus$, raising a tide on a body (the Moon) 
of radius $R_{\leftmoon}$ can be expanded in
powers of $r$, the distance between the two bodies, and spherical harmonics  \citep{kaula64,peale78}.
Neglecting the  angular dependence,
the quadrupole term $V_2 \sim \frac{GM_\oplus}{r}  \left( {R_{\leftmoon}/ r} \right)^2$
where $G$ is the gravitational constant.
The next  term in the expansion is the octupole term with
$ V_3 \sim \frac{GM_\oplus}{r}   \left( {R_{\leftmoon}/ r} \right)^3$. 
The relevant octupole spherical harmonic is lopsided with strength on the
Moon's near side opposite in sign to that on the far side.   The octupole term arises because
gravitational attraction from a perturber is stronger on the near side.  The closer the perturber
is to the body, the worse the quadrupole tidal approximation and the more important are higher order
moment terms such as  the octupole.

As the radius of a
tidally heated body is usually small compared to its semi-major axis (for example the ratio of 
semi-major axis to radius for Io is 230 and that for Enceladus is 944),
the octupole and higher order moments are usually neglected in tidal
heating computations.   An exception is 
Phobos,  at only 2.76 Mars radii away from Mars,  higher order terms are considered in
modeling its tidal evolution (and drift rate in semi-major axis) due to tidal dissipation in Mars 
(rather than Phobos) \citep{bills05}.
If the Earth's tidal dissipation parameter $Q_\oplus$ is large, $\sim 10^4 $ to $10^6$ 
(corresponding to weak dissipation), while it hosted a magma
ocean, as proposed by \citet{zahnle15}, then lunar crustal growth could take place
while the Moon's semi-major axis was only $a_{\leftmoon} \lesssim 6 R_\oplus \sim 20 R_{\leftmoon}$.    
We explore the possibility that 
the octupole tidal perturbation could influence the tidal heating pattern during the era
of lunar crustal growth, and cause a near/far side asymmetry.

Taking into account the body's spin rate and the orbit variations, the tidal potential perturbation is 
expanded in a Fourier series where each term depends on a different frequency 
(e.g., equation 1 by \citealt{kaula64} and \citealt{peale78}).
An elastic  body is deformed by the tidal perturbation giving stress and strain tensors that are similarly
expanded as a Fourier series with the same frequency spectrum.
The strain tensor $\epsilon_{ij}$ is computed from the displacement vector and with strain
tensor components and displacement vector
proportional to the potential perturbation \citep{love44,peale78}.
For  the quadrupole potential tidal perturbation  the strain tensor components
\begin{equation}
\epsilon_{ij,2} (\theta,\phi) \sim k_2 \left( \frac{M_\oplus}{M_{\leftmoon}}\right) \left(\frac{R_{\leftmoon}}{a_{\leftmoon}}\right)^3 .
\end{equation}
(equations 10--15 by \citealt{peale78}).  Here $k_2$ is a unit-less Love number.
 Because these components are proportional to the potential term,
 the octupole term in the tidal perturbation  induces a similar strain, 
but proportional to one higher power of $R_{\leftmoon}/a_{\leftmoon}$;
\begin{equation}
\epsilon_{ij,3} (\theta,\phi) \sim k_3 \left( \frac{M_{\oplus}}{M_{\leftmoon}} \right) \left(\frac{R_{\leftmoon}}{a_{\leftmoon}}\right)^4,
\end{equation}
where $k_3$ is the Love number for the octupole perturbation term.
The total strain is a sum of multipole terms and  is sensitive to the angular dependence  
of each component (on spherical coordinate angles $\theta,\phi$).

Tidal dissipation, giving  heat per unit volume, $h$, is computed from the average over time of
stress times strain rate 
\begin{equation}
h_{\rm tidal} =  \sum_{ij} \langle \sigma_{ij} \dot \epsilon_{ij} \rangle = \sum_{ij} \langle \mu \epsilon_{ij} \dot \epsilon_{ij} \rangle,  \label{eqn:hW}
\end{equation}
where both the stress and strain rate consist of a sum of terms
(see equation 20 by \citealt{peale78}). Here the stress tensor is $\sigma_{ij}$, 
the strain rate is $\dot \epsilon_{ij}$, and the stress and strain are related by the shear modulus $\mu$
which can be a complex function, to take into account phase lag due to viscoelastic relaxation. 

The strain rate depends on the strain times the tidal perturbation frequency.
If the octupole term in the expansion is considered alone, we would expect the tidal heating
rate to be $(R_{\leftmoon}/a_{\leftmoon})^2$ times weaker than the classical tidal heating
rate computed from the quadrupole term alone.
However, the product of stress times strain rate contains equivalent frequency terms
that are proportional to the product $V_2 V_3$ or $\epsilon_{ij,2} \epsilon_{ij,3}$ and so are only a factor  
$(R_{\leftmoon}/a_{\leftmoon})$
times weaker than the quadrupole tidal heating rate.
The quadrupole heating pattern is symmetrical, heating the near and far sides of the Moon equivalently.
However, inclusion of the octupole potential term in a computation of the heating rate
would give a difference between near and far sides of order  $(R_{\leftmoon}/a_{\leftmoon})$.
This would give a 5--10\% asymmetry in the tidal heating rate during the era of lunar crustal growth.
An even larger asymmetry might be present as the ratio of maximum to minimum in the distribution of tidal heat 
could be  larger than 1.
With the quadrupole alone, the heat flux (heat per unit area) on the surface can vary 
by a factor of 2 (see for example, \citealt{beuthe13}).
Asymmetric tidal heating could be more important than a few percent difference in heat flux through
the lunar crust due to Earth-shine (as we estimated in section \ref{sec:timeline}). 

\subsection{Outline}


The tidal dissipation rate inside the Moon has predominantly been predicted using
a semi-analytical expansion method (e.g., \citealt{peale78,beuthe13,beuthe15}).
We use a numerical simulation technique instead. 
Because of their simplicity and speed, compared to more computationally intensive grid-based or finite element methods,
 mass-spring computations are a rough but straightforward method for simulating deformable bodies.
By including spring damping forces they can model viscoelastic tidal deformation.  Because
all forces are applied between pairs of particles,  angular momentum conservation is ensured.
Extremely small strains can be precisely measured between two mass nodes so tidal deformation is measurable.    
We have previously used a mass-spring model to study tidal encounters  \citep{quillen16_crust}, 
measure tidal spin down for spherical bodies over a range
of viscoelastic relaxation timescales  \citep{frouard16},  spin-down of triaxial bodies 
spinning about a principal body axis aligned with the orbit normal \citep{quillen16_haumea},
obliquity evolution of minor satellites in the Pluto-Charon binary system \citep{quillen17_pluto},
excitation of normal modes and seismic waves on a non-spherical body \citep{quillen19_bennu}
and dissipation in ellipsoids undergoing non-principal axis rotation \citep{quillen19_wobble}.

In this study we use the mass spring models to track the distribution of tidally generated heat
in a soft  nearly spherical body, mimicking the Moon.  
The Moon is resolved with nodes and  a spring network,
and is in an eccentric orbit about its  tidal perturber, here the Earth, that is modeled as a point mass.
We measure the heat distribution from the dissipation rates in the springs.  
Tidal heating computations often assume spherically symmetric shell models for internal material properties such as
density,  composition, rigidity and viscosity 
(e.g.,  \citealt{peale78,ojakangas89,tobie05,wahr06,nimmo07,wahr09,beuthe13,beuthe15}). 
Our mass spring model simulations are not restricted to spherical symmetry.

We integrate the tidally generated heat  giving a heat flux pattern
as a function of latitude and longitude.
Our simulated Moon has a dissipating shell, approximating a thin  crust,
on a soft interior, mimicking the lunar magma ocean. 
Our simulations are described in section \ref{sec:sims}.
In section \ref{sec:heat}
we compare tidally generated heat distributions for a series of simulations with uniform
crustal shell thickness
that have different orbital semi-major axes.  We examine the tidal heat distribution of a simulation with
 a crustal shell that has  varying thickness.
Models for asymmetric lunar crust growth are  discussed in section \ref{sec:asymm}.
A summary and discussion follows in section \ref{sec:sum}.

\section{Simulations}
\label{sec:sims}

\begin{table*}[ht!]
\begin{center}
\caption{\large  Moon Parameters and values used for heating models \label{tab:moon}}
\begin{tabular}{@{}lllllll}
\hline
Mean radius                    &  $R_{\leftmoon}$ & 1737.1 km  \\
Mass                               & $M_{\leftmoon}$ & $7.342 \times 10^{22}$ kg \\
Mean density                   & $\rho_{\leftmoon}$ & 3344 kg\ m$^{-3}$\\
Surface gravity                 & $g_{\leftmoon}$ & 1.62 m\ s$^{-2}$ \\
Equatorial surface temperature      & $T_s$ & 255$^\circ$ K \\
Magma ocean temperature           & $T_b$ & 1175$^\circ$ C = $1450^\circ$ K \\
\hline
Crustal thermal conductivity  & $K_T$     & 1.8  W~m$^{-1}\  {\rm K}^{-1}$\\
Parameter for viscosity activation  & $\gamma_Q (T_b - T_s) $ & 44 \\
Latent heat of fusion in magma ocean & $L_{\rm fusion}$  & $4 \times 10^5$ J/kg\\
Density of plagioclase & $\rho_{\rm plag}$ & 2670 kg~m$^{-3}$ \\
Proportion of plagioclase in ocean & $f_{\rm plag}$  & 0.2 \\
\hline 
\end{tabular}
{\\ Notes: In the second half of the table, the nominal value of $K_T$ 
is that used by \citet{garrickbethell10}.  The parameter for viscosity activity is discussed
in section \ref{sec:asymm}.  The proportion $f_{\rm plag}$ is that used by \citet{snyder92,warren86}.
The latent heat  $L_{\rm fusion} $ is the nominal value used by  \citet{tian17}.
}
\end{center}
\end{table*}

\begin{table}[ht!]
\begin{center}
\caption{\large   Nomenclature \label{tab:nomen}}
\begin{tabular}{@{}lllllll}
\hline
Minimum interparticle spacing & $d_I$   \\
Maximum spring length           & $d_S$  \\
\hline
Number of nodes                     & $N$ \\
Mass of a node                        & $m_i$ \\
\hline
Number of springs                   & $\NS$ \\
Length of a spring                   & $L_{ij}$ \\
Rest length of a spring            & $L_{ij,0}$ \\
Spring constant (elastic)            & $k_{ij}$ \\
Spring damping parameter        & $\gamma_{ij}$ \\
Strain of a spring                       & $\epsilon_{ij}$ \\
Energy dissipation rate of a spring    & $\dot E_{ij}$ \\
\hline
Simulation time step & $dt$ & \\
Initial damping time &  $t_{\rm damp}$ & \\
\hline
Total mass  of resolved body    & $M$ or $M_{\leftmoon}$ \\
Radius of resolved body    & $R$ or $R_{\leftmoon}$\\
Angular spin rate                      & $\Omega$ \\
Mass of perturber                    & $M_*$ or $M_\oplus$ \\
Orbital Eccentricity                  & $e_o$ \\
Orbital semi-major axis           & $a_o$ \\
Orbital mean motion              & $n_o$ \\
\hline
Poisson ratio & $\nu$ \\
Shear modulus & $\mu$\\
Viscosity & $\eta $\\
Viscoelastic relaxation time & $\tau_{\rm relax}$ \\
\hline
Strain tensor & $\epsilon_{ij}$ \\
Stress tensor & $\sigma_{ij}$ \\
Temperature & $T$ \\
Heating rate per unit volume & $h$ \\
Heat flux or heating rate per unit area & $q$ \\
Crustal or shell thickness & $H$ \\
Thermal conductivity  & $ K_T$ \\
Latitude  & $ \vartheta$ \\
Longitude & $\phi$ \\
Coordinate for depth in crust & $z$ \\
Tidal frequency & $\sigma_t$ \\
Hard/Soft Shell coupling parameter & $R_{HS}$ \\
\hline
\end{tabular}
\end{center}
{Notes:  A subscript $i$ is put on a quantity at a mass node with index $i$.   Subscript $ij$ refers to a spring
connecting mass node $i$ to node $j$.  Subscripts $ij$ for the stress and strain tensors refer to coordinate
directions.  
}
\end{table}

\begin{table}[ht!]
\begin{center}
\caption{\large  Common Simulation Parameters \label{tab:common}}
\begin{tabular}{@{}lllllll}
\hline
Minimum interparticle distance & $d_I$ &  0.11 \\
Max spring length to min particle spacing  & $d_S/d_I$  & 2.3 \\
\hline
Number of nodes                     & $N$  & $ 2085$ \\
Number of springs  per node   & $\NS/N$ & $ 13$ \\
Spring constant in interior         & $k_{I}$ & $0.03$\\
Spring constant in shell            & $k_{S}$ & $0.16$ \\
Damping parameter   in interior      & $\gamma_I$&  $0.01$\\
Damping parameter   in shell         & $\gamma_S$& 1\\
Young's modulus of interior &  $E_I$ & 1.3 \\
Young's modulus of shell    &  $E_S$ & 7  \\
Viscoelastic relaxation time in shell & $\tau_{S, {\rm relax}}$ & 0.0015  \\
Shell boundary equatorial radius   & $R_S$    & 0.7 \\
\hline
Initial body spin                  & $\Omega$ & 0.5  \\
Orbital eccentricity              & $e_o$  & 0.2 \\
Time step & $dt$ & 0.005 \\
Damping time &  $t_{\rm damp}$ & 20  \\
Orbital period & $2 \pi/n_o $  & $4 \pi$ \\
\hline
\end{tabular}
\end{center}
{Notes: Due to variations in the generation of the random spring network, the number of springs per node
and number of nodes  varies from simulation to simulation; 
$N$ varies by about $\pm 5$ and $\NS$ varies by about $\pm 120$.
Additional simulation parameters are listed in Table \ref{tab:series}.
}
\end{table}

\begin{table}[ht!]
\begin{center}
\caption{\large  Simulation Series \label{tab:series}}
\begin{tabular}{@{}lllllll}
\hline
Simulation                                              & M1 & M2 & M3 & M4 & O2 \\
\hline
$a_o/R$  & 7 & 15 & 30 & 60 & 15 \\
$a_o/R_\oplus = 3.7 \times a_o/R$ & 2 & 4 & 8 & 16  & 4 \\
Mass ratio  $M_*/M$   & 81 & 813 & 6500 &$5 \times 10^4$ & 813 \\
\hline
\end{tabular}
\end{center}
{Notes:  $a_o/R$ is the ratio of orbital semi-major axis to lunar radius.
3.7 times this represents the the semi-major axis in units of Earth radius.
The M1--M4 simulations have spherical symmetry.
The O2 simulation has an uneven crustal shell thickness.
Additional simulation parameters are listed in Table \ref{tab:common}.
}
\end{table}

To simulate tidal viscoelastic response of non-spherical bodies
we use the mass-spring model \citep{quillen16_crust,frouard16,quillen16_haumea,quillen17_pluto,quillen19_bennu}
that is built on the modular N-body code \texttt{rebound}  \citep{rebound}.
An elastic solid is approximated as a collection of $N$ mass nodes that are
connected by a network of $\NS$ massless springs.
Springs between mass nodes are damped and
so the spring network approximates the behavior of a Kelvin-Voigt viscoelastic
solid with Poisson ratio of 1/4 \citep{kot15}.  

The mass particles in the resolved spinning body are subjected to three types of forces:
the gravitational forces acting on every pair of particles in the body and from external masses,
 and the elastic and damping spring forces
acting only between sufficiently close particle pairs with previously identified springs connecting them. 
When a large number of particles is used to resolve the spinning body, the mass-spring model behaves
like an isotropic continuum elastic solid \citep{kot15,kot15b}, including its ability to exhibit normal
mode oscillations and transmit seismic waves \citep{quillen19_bennu}. The Kelvin-Voigt model arises because 
spring forces are exerted in parallel with damping.  The spring forces keep a self-gravitating body
from collapsing.  A Maxwell viscoelastic model, with spring and damper in series between pairs of massive particles, 
would collapse due to self-gravity.
Symbols used to describe our simulations are summarized in Table \ref{tab:nomen}.

The code has been checked by comparing simulated and predicted tidal heating rates for tidal spin down 
\citep{frouard16} and by comparing simulated and predicted energy dissipation rates (wobble damping) for
ellipsoids undergoing non-principal axis rotation \citep{quillen19_wobble}.
The code matches analytical predictions within 30\%, even at low stress and strain.
We attribute the discrepancies in the tidal spin down comparison to the neglect of bulk viscosity 
in the analytical calculation, not to a problem in the simulation technique.
The differences between measured and predicted dissipation rates for wobble damping are the same
size as differences between different two different analytical predictions 
and match these predictions over 4 orders of magnitude variation in the energy dissipation rate \citep{quillen19_wobble}. 
  We have also compared
predictions using a quality factor (constant $Q$) dissipation model, a Maxwell model and our simulations 
which obey a Kelvin-Voigt viscoelastic model \citep{quillen19_wobble}.
Sensitivity to numbers of particles simulated, spring network or lattice and particle generation technique is discussed by \citet{quillen16_haumea}.
Normal mode frequencies and seismic wave propagation speeds were checked  for a near-spherical shape model 
by \citet{quillen19_bennu}.  We have also used this code to study spin-orbit resonance capture and crossing 
in multiple body systems \citep{quillen17_pluto}.

\subsection{Sizes and units}

Physical quantities are summarized in Table \ref{tab:moon}.
Our simulations work with mass in units of $M = M_{\leftmoon}$,  
the mass of the resolved spinning body, here that of the Moon,
and distances in units of volumetric radius, $R = R_{\rm vol}$, 
the radius of a spherical body with the same volume. 
Here we take $R_{\rm vol} = R_{\leftmoon}$, the Moon's mean equatorial radius.
The ratio $R_\oplus/R_{\leftmoon} = 3.67$ and we use this factor to compute
the ratio of  semi-major axis  to  Earth radius, $a_{\leftmoon}/R_\oplus$. 

For our mass-spring simulations  it is convenient to work with time in units of a gravitational timescale
\begin{align}
t_{\rm grav} &\equiv \sqrt{\frac{R^3}{GM}} = \sqrt{ \frac{3}{4 \pi G \rho}   } \nonumber \\
&= 1034\ {\rm seconds}  \left( \frac{\rho_{\leftmoon} }{\rho} \right)^\frac{1}{2}  
\label{eqn:tgrav}
\end{align}
where on the last line we have used the mean density of the Moon, $\rho_{\leftmoon }$
(listed in Table \ref{tab:moon}). 
It is  helpful to define a unit of energy density
\begin{align}
e_{\rm grav} & = \frac{GM^2}{R^4} \nonumber \\
&= 39.5\ {\rm GPa} \left( \frac{M}{M_{\leftmoon } } \right)^2 \left( \frac{R_{\leftmoon} }{R}\right)^4
\label{eqn:eg}
\end{align}
where we are using the Moon's mass $M_{\leftmoon}$ and mean radius $R_{\leftmoon}$. 
Pressure, energy density and elastic moduli are given in units of  $e_{\rm grav}$.
The velocity of a massless particle in a circular orbit just grazing the surface of the body
is $1$, and the period of this orbit is $2 \pi$.

\subsection{Initial conditions}

As done previously \citep{frouard16}, we consider two masses in orbit.  
One mass is the spinning body resolved with masses and springs.
The other body, the tidal perturber,  is  a point mass with mass $M_*$.
The resolved body is the Moon, and the point mass perturber is the Earth.
For our random spring model, particle or node positions for the resolved body are drawn from an isotropic uniform distribution but only accepted into the spring network if they are within the surface bounding a triaxial ellipsoid, 
 $x^2/a^2 + y^2/b^2 + z^2/c^2 = 1$, and if they are more distant than $d_I$ from every other previously generated particle. 
 Particles nodes that are not inside the ellipsoid are deleted.
 Here $a,b,c$ are the body's semi-major axes.  
For modeling the Moon we generate a spherical body with $a=b=c=1$.
Once the particle positions have been generated, every pair of particles within $d_S$ of each 
other are connected with a single spring.    
Springs are initiated at their rest lengths.  
Because of self-gravity the body is not exactly in hydrostatic equilibrium at the beginning
of the simulation and the body initially vibrates.
We apply additional damping at the beginning of the simulation
for a time $t_{\rm damp}$ to remove these vibrations.  Only after $t_{\rm damp}$ do we process simulation
outputs for measurements of dissipation.     After this time spring parameters are not varied.

The spherical resolved body is initially
spinning at rate $\Omega$  
with spin axis aligned with the orbital angular momentum axis, and in an 
eccentric orbit about the point mass.
As we are modeling a tidally locked Moon
we start with spin $\Omega = n_o$ with $n_o = \sqrt{G(M_* + M)/a_o^3}$, the orbital mean motion.
Here  $a_o$ and $e_o$ are   the orbital semi-major axis and eccentricity. 

\subsection{Energy Dissipation in the Mass-Spring model}

We describe how we use the mass spring model simulations to  track the energy dissipation
 in each spring.
We consider a spring between two particles $i,j$ with coordinate positions ${\bf x}_i$, ${\bf x}_j$.  
The vector between the particles
${\bf x}_i - {\bf x}_j$ gives a spring length $L_{ij} = |{\bf x}_i - {\bf x}_j| $ that we compare with 
the spring rest length $L_{ij,0}$.  The spring strain is 
\begin{equation}
\epsilon_{ij} = (L_{ij} - L_{ij,0})/L_{ij,0}.
\end{equation}
The strain rate of a spring with length $L_{ij}$ is
\begin{equation}
 \dot \epsilon_{ij} = \frac{\dot L_{ij}}{L_{ij,0}} =
 	 \frac{1}{L_{ij} L_{ij,0}} ({\bf x}_i - {\bf x_j}) \cdot ({\bf v}_i -  {\bf v}_j) .
 \end{equation}

The elastic force on particle $i$ by the spring connecting $i,j$ is
\begin{equation}
{\bf F}_i^{\rm elastic} = -k_{ij} (L_{ij} - L_{ij,0}) \hat {\textbf{n}}_{ij}  \label{eqn:elasticforce}
\end{equation}
with $k_{ij}$ the spring constant and the unit vector $\hat {\bf n}_{ij} = ( {\bf x}_i - {\bf x}_j )/ L_{ij}$.
Our damping force on particle $i$ is proportional to  the strain rate
\begin{equation}
 {\bf F}_{i}^{\rm damping} = - \gamma_{ij} \dot \epsilon_{ij} L_{ij,0} m_{ij} \hat{\textbf{n}}_{ij} \label{eqn:dampforce}
 \end{equation}
with damping coefficient $\gamma_{ij}$  that is equivalent to the inverse of a damping  time scale. 
 Here $m_{ij}$ is the reduced mass $m_{ij} = m_i m_j /(m_i + m_j)$.
The elastic and damping forces on particle $j$ have the opposite sign as in equations \ref{eqn:elasticforce} and \ref{eqn:dampforce}.

Taking the sum of damping force times velocity for both masses we find 
the energy dissipation rate in a single spring connecting node $i$ to $j$ is
\begin{align}
\dot E_{ij} &= {\bf F}_{i}^{\rm damping} \cdot ({\bf v}_i - {\bf v_j}) \nonumber \\
&= - \frac{\gamma_{ij} m_{ij}}{L_{ij}^2} \left[ ({\bf x}_i - {\bf x_j}) \cdot ({\bf v}_i -  {\bf v}_j)  \right]^2 \nonumber \\
&= -\gamma_{ij} m_{ij} L_{ij,0}^2 \dot \epsilon_{ij}^2 \nonumber\\
&= -\gamma_{ij} m_{ij} \dot L_{ij}^2 . \label{eqn:dotE_ij}
\end{align}

At each time step of a simulation, we record $\dot E_{ij}$ for every spring.
In a frame rotating with the body we average the tidal power per unit volume taking values of $\dot E_{ij}$ for
each spring  over a series of time steps.  
The mid-point of the spring is used to identify the location of generated
heat.
By counting the number of springs per unit volume and measuring  their dissipation rate
we create 3-dimensional internal maps of the tidal heating rate per unit volume.

\subsection{Viscous relaxation time}

We summarize the relation between the springs in the model and the approximated
continuum material (following \citealt{kot15,quillen16_crust,frouard16}).
Taking the continuum limit,
the static Young's modulus for the random mass spring model is computed from a sum over  springs
within a volume $V$ \citep{kot15}
\begin{equation}
E = \frac{1}{6V} \sum_{i,j>i} k_{ij} L_{ij,0}^2,
\end{equation}
where $L_{ij,0}$ is the rest length of a spring connecting node $i$ to node $j$ and  $k_{ij}$
is the spring constant of that spring.
This equation expresses how a network with stronger springs and more springs per unit volume would give a stronger material.
The shear modulus is 
\begin{equation}
\mu = \frac{E}{2(1 + \nu)} = \frac{E}{2.5},
\end{equation}
where the Poisson ratio $\nu=1/4$ for the random spring model.  

The shear viscosity is similarly computed \citep{quillen16_crust,frouard16} from a sum over
springs within a volume
\begin{equation}
\eta = \frac{1}{2(1+\nu) } \frac{1}{6V} \sum_{i,j>i} \gamma_{ij} m_{ij} L_{ij,0}^2 ,  \label{eqn:eta}
\end{equation}
giving a relaxation time 
\begin{equation}
\tau_{\rm relax} = \frac{\eta}{\mu}.
\end{equation}
Equation \ref{eqn:eta} expresses how a network with more heavily damped springs mimic a higher viscosity material.
With nodes of the same mass, $m$, and springs with the same damping parameter $\gamma$
and spring constant $k$, the viscoelastic relaxation time
\begin{equation}
\tau_{\rm relax} = \frac{\gamma m}{2 k},
\end{equation}
where the factor 2 arises because we have used the reduced mass in equation \ref{eqn:dampforce}.

We use a two layer model with properties in the interior denoted with subscript $I$ and those
in the outer shell denoted with subscript $S$, 
giving moduli $E_S, E_I$, viscosities $\eta_S, \eta_I$ and relaxation timescales
$\tau_{S, {\rm relax}}, \tau_{I, {\rm relax}}$. 

A body in spin synchronous or tidally locked state has  tidal frequency $\sigma_t \approx n_o$ the mean motion.
The tidal frequency in units of the relaxation time $\bar \chi = | \sigma_t| \tau_{\rm relax}$. 
For $\bar\chi <1$, the quality function, 
giving the torque and integrated energy dissipation,
is $\propto \bar \chi$ (see section 2.3  by \citealt{frouard16}) .
In our simulations we chose damping parameter 
$\gamma_S$ so as to remain in the linear regime
and approximately giving a constant time lag 
tidal dissipation model.   
The tidal quality factor as a function of frequency 
for the Kelvin-Voigt rheology is qualitatively similar to that of the Maxwell rheology  \citep{frouard16}. Both 
rheologies have quality factor that is 
 linear at low frequencies and both reach a peak near $\bar \chi \sim 1$ \citep{frouard16}.
Because we remain in the linear regime, our simulated material should give a tidal
heat distribution similar to that predicted by a Maxwell model and with the same frequency dependence.

\begin{figure}
 \centering
    \includegraphics[width=3in]{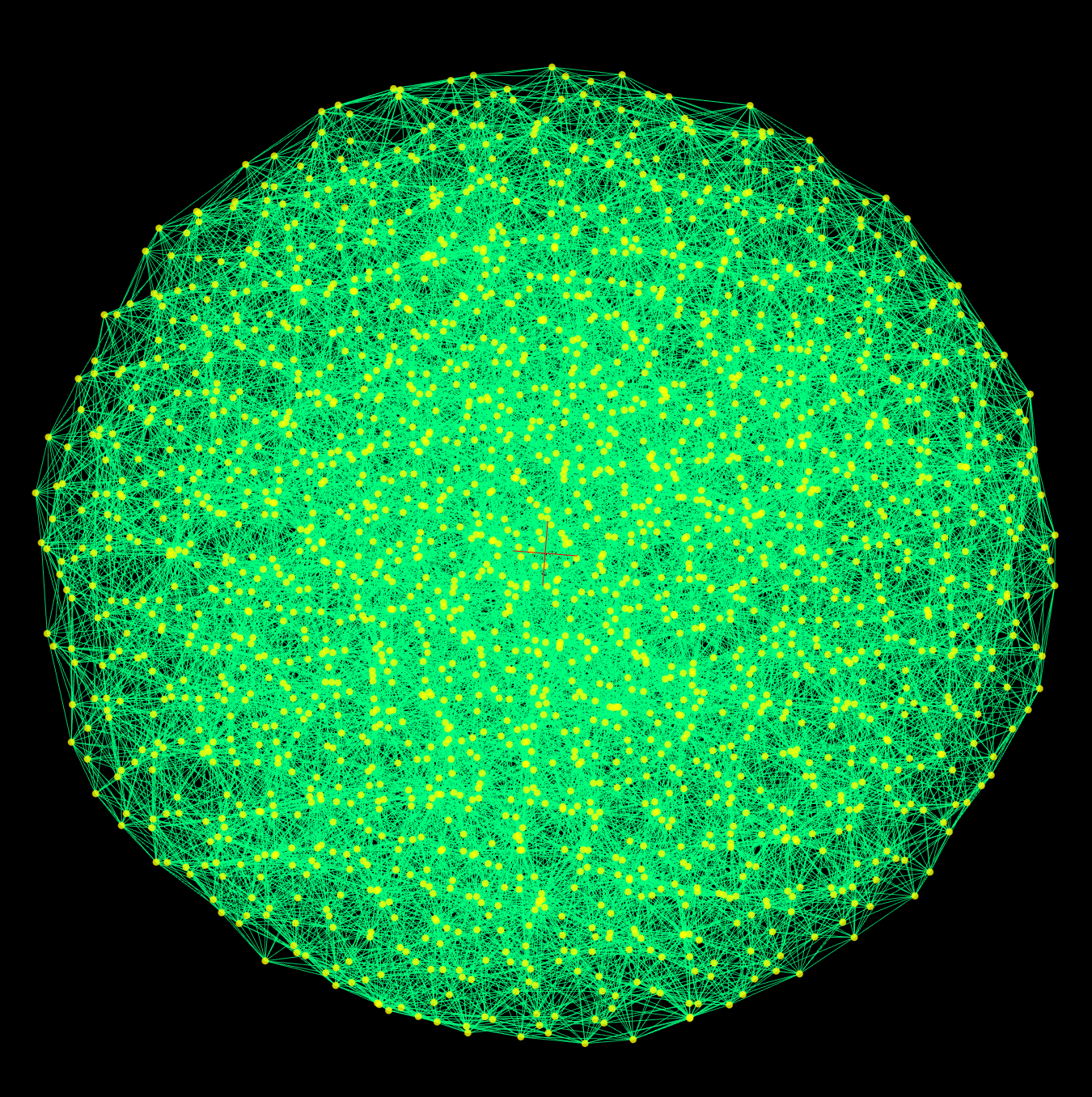} 
 \caption{We show the spring network from a visualization of one of the simulations.
 Green lines are springs and yellow dots are the mass nodes.
 \label{fig:network}}
\end{figure}

\subsection{Simulation parameters}

Common simulation parameters are listed in Table \ref{tab:common}.  A simulation visualization (a screen shot) 
showing the spring
network from one of the simulation is shown in Figure \ref{fig:network}.
The resolved body is generated with mass nodes all identical in mass and is approximately 
uniform in mass density.
Springs are generated with a two layer model.  We use an ellipsoidal boundary to separate the crustal shell from
the soft interior.
Springs with midpoint inside this  boundary  
 have spring constants and damping parameter, $k_I, \gamma_I$, and those outside it, in the stiffer shell,
are $k_S, \gamma_S$.
The interior is adjusted to have a low damping coefficient, $\gamma_I$ giving it a low dissipation rate.
To mimic crustal heating we set the damping rate in the shell springs higher than that in the interior, $\gamma_S > \gamma_I$.
A low Young's modulus $E_I \sim 1.3$ is chosen for the interior to mimic  a magma ocean.
Our simulations do not model fluids  and because the springs  
 keep the body from collapsing due to self-gravity, we cannot simulate a material
with Young's modulus less than 1.  A higher Young's modulus is chosen for the shell, $E_S = 7$,
to mimic the behavior of a stiffer but deformable elastic crust.
The Young's modulus for our simulation is in gravitational units, but for the Moon $e_g = 40$ GPa
(equation \ref{eqn:eg})
so a modulus of $E_S \sim$ 2 in gravitational units is similar to that of rock (80 GPa).
Our simulations have a shell that is significantly harder than rock so that we can mimic the behavior of a hard shell
on top of a softer interior.

We require accurate measurement of gravitational forces so that small variations in
stresses and strains in the springs can be computed.   As a result, we compute the gravitational forces
between all pairs of particles during every time step and the 
the number of computations depends on the square of the number of mass nodes, $N^2$
\citep{quillen16_crust,frouard16,quillen16_haumea}.
The mean distance between mass nodes depends on $N^{1/3}$, so a large increase
in the number of nodes is required to increase the resolution of the simulation.  The simulation time-step $dt$ depends
on the time for elastic waves to travel between mass nodes and so also scales with $N^{1/3}$. 
Our simulations are not yet parallelized, 
so we cannot yet extend them to significantly larger numbers of nodes than a few thousand
for simulations that are carried out for a few hundred orbital periods.   The choice of run time 
depends upon  which quantities we want to measure 
and their sensitivity to various oscillation frequencies (such as precession and libration).

We show five simulations, listed in Table \ref{tab:series} and with additional parameters in Table \ref{tab:common}.  
Each has initial body spin $\Omega=0.5$ in gravitational units
and orbital mean motion $n_o= \Omega$, so the body starts in tidal lock (a spin synchronous state).
The  simulations each have a different semi-major axis.  
So that all simulations have the same body spin and the same orbital period, the mass of the perturber $M_*$
is adjusted so that the mean motion $n_o = \sqrt{\frac{M_* + 1}{a_o^3}} = \Omega$, (here $M_*$ is in units of $M$) giving larger perturber
masses at larger orbital semi-major axes.    The tidal heating rate is proportional to the mass of the perturber,
however, the distribution of heat is not dependent on the perturber mass. 
We opted to adjust the perturber mass rather than the orbital frequency so we could fix the 
 tidal frequency and viscoelastic relaxation time.

By varying the ratio of semi-major axis to body radius
we can look at how proximity affects the tidal heating distribution.
The orbital eccentricities for our simulations is 0.2.  The tidal heating rate increases with eccentricity,
however  the predicted pattern of tidal heating as a function of latitude and longitude is not  dependent on it \citep{peale78}.
We checked that the normalized heating distributions in our simulations are insensitive  to orbital eccentricity.

\begin{figure*}
 \centering
    \includegraphics[width=4in]{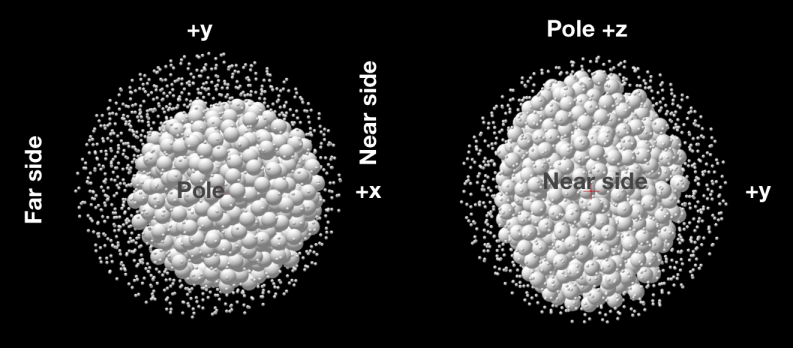} 
 \caption{We show the lopsided shell for the O2 simulation.    
 We have inflated the rendered size of core mass nodes,
compared to nodes in the shell.  
 The left image is a simulation visualization looking down from above a pole.
 The right image shows a simulation snap show  looking in the equatorial plane.
 The shell in the O2 simulation is thinner on the poles and on the side nearest the perturber.
 The mass nodes are rendered as spheres but they behave as point masses. 
 \label{fig:o2_lop}}
 \end{figure*}

The simulations M1, M2, M3, and M4 have spherical, uniform thickness crustal shells with inner radius 
$R_S$ but differing orbital semi-major axes.  
The simulation
O2 has an asymmetric crustal shell and its orbital semi-major axis is the same as the M2 simulation.  
The shell boundary for this simulation is described with an ellipsoid 
\begin{equation}
 \left(\frac{x-x_S}{R_S} \right)^2 + \left(\frac{y-y_S}{R_S} \right)^2 + \left(\frac{z-z_S}{c_s} \right)^2 = 1 \label{eqn:shell_eps}
\end{equation}
with positive $x$ initially toward the perturber and the $z$ axis pointing toward a pole.
The O2 simulation has shell boundary equatorial radius $R_s = 0.7$, 
polar semi-axis $c_s = 1.2 R_S$ and offset $(x_S,y_S,z_S) = (0.15,0,0)$ so that the
shell is thinner at the poles and on the near side.   By enlarging the rendered radii of the interior mass nodes we show in 
Figure \ref{fig:o2_lop} the shape of the shell with polar and equatorial views.

Due to the uneven particle distribution in the resolved body the initial body spin vector differs slightly
from the desired value and the body spin axis is not exactly aligned with its principal body axes.
Our simulations are not initially exactly at the lowest spin energy state and the body can oscillate about
the spin synchronous state.   This oscillation is sometimes called free libration.
We don't yet know how to set the initial conditions accurately enough to minimize the free libration amplitude.
The  time for an amplitude in free libration  to die away is long as the libration frequency
 is slow for a nearly homogeneous sphere in a near 
spin synchronous state.
We attempt to set the initial spin so that the initial obliquity is zero, however, the uneven particle
distribution causes the spin vector to be slightly misaligned with principal body axes giving  slow body precession.    
Longer simulations often showed true polar wander. 
We mitigate these problems with a high initial damping rate in the shell and by only running the
simulations for 15 orbital periods afterward. 

\begin{figure*}
\centering
    \includegraphics[width=5in]{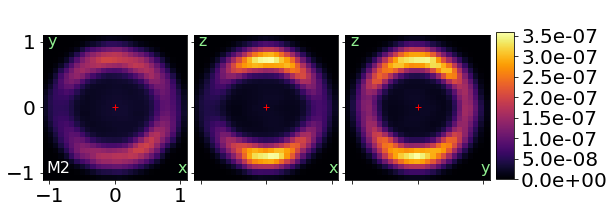} \\
    \includegraphics[width=5in]{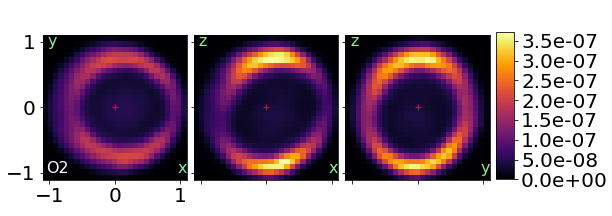} 
 \caption{
Internal heat maps in three planes that bisect the body for the M2  simulation (top row of panels) and the O2 simulation (bottom row of panels).  The shell has a uniform thickness in the M2 simulation. The O2 simulation has a thinner shell at the poles and on the side nearest the perturber (at positive $x$).
The heating rate is measured from the dissipation rate in springs in a mass spring model simulation.
Energy dissipation rates of springs within a distance $\sim 0.2$ of the bisecting plane are summed to produce these map which
shows the heating rate per unit volume.
The $x$ axis is toward the Earth, and the $z$ axis parallel to the orbital angular momentum and giving the poles.
The leftmost panels shows the heating rate per unit volume in the $xy$ or equatorial plane.
The middle panels shows the heating rate in the $xz$ plane, containing the poles and near and far sides.
The rightmost panel shows the heating rate in the $yz$ plane.
The pattern of heat is primarily polar, as expected for a shell model for a librating tidally locked body
in eccentric orbit \citep{peale78}.  Heating is stronger at the base of the shell and weak in the interior
because the springs damp more strongly in the shell than the interior.  
 \label{fig:heat}}
 \end{figure*}

\section{The spatial distribution of tidally generated heat}
\label{sec:heat}

The distribution of tidally generated heat per unit volume
on three planes that bisect the body are shown in Figure \ref{fig:heat} for simulations M2 and O2 with parameters
listed in Tables \ref{tab:common} and \ref{tab:series}.
In Figure \ref{fig:heat} the positive $x$ axis points toward the perturber (the Earth) so $x=1$ is the Moon's near side.
The $z$ axis is aligned with spin axis and is perpendicular to the orbital plane, giving the poles.
Positive $y$ corresponds to trailing side in the orbital plane (away from the direction of rotation).
The Moon's equator is in the $xy$ plane.
Units of energy dissipation per unit volume, $\dot e$, are gravitational or  $e_g/(R^3 t_{grav})$.
The heating rate is measured from a sum of dissipation rates in the individual springs, but  only springs
near the bisecting plane and with midpoints within the pixel are summed.    To decrease the sensitivity of the heat maps to node positions,
we weight the sum with a Gaussian weight that depends
on distance from the bisecting plane and has a dispersion of 0.2 in units of radius.   The resulting
weighted heat distribution for  three bisecting planes and for two simulations  is  shown in Figure \ref{fig:heat}.

The heat distributions in the center and right panels of Figure \ref{fig:heat} show that the 
heat per unit volume is highest at the poles.  This is expected for a tidally locked body
that is librating  in an eccentric orbit \citep{peale78,ojakangas89,tobie05,beuthe13,beuthe15}.
Tidal heating is low in the interior because the springs there are only weakly damped.
Because the tidal stress is higher at the base of the shell, the heating rate is higher at the shell
base than near the surface; this too is predicted from semi-analytical tidal heating models 
\citep{peale78,garrickbethell10,tobie05,beuthe13,beuthe15}.

In the O2 simulation (shown in the bottom row of panels in Figure \ref{fig:heat}), the shell is thinner at the poles
and on the near side.  The thickness of the far side shell is evident
in the  heat distribution in the equatorial plane shown in the leftmost panel. 
The heating pattern is similar to the M2 simulation even though the heat distribution appears oval
for the right two panels because the core is extended in the $z$ direction.
 The heat distribution is a tilted oval because
the body has not remained oriented so that it is aligned with the orbit and with the direction of the perturber.

The panels on the left, show heating in the equatorial plane.  In the M2  simulations, there is a weak asymmetry,
with the near side (positive $x$) somewhat hotter than the far side.  The asymmetry can also be seen in the middle panel
as the polar heating pattern is slightly warmer on the near side.  
 We attribute the asymmetry to the proximity of the tidal
perturber, making the octupole moment in the gravitational potential from the perturber 
strong enough to cause asymmetry in the heating rate.   
We discuss this asymmetry in more detail below.

\begin{figure*}
  \vspace{-0.4in}
$$
\begin{array}{cc}
 \includegraphics[width=2.7in, trim=14 4 30 10, clip]{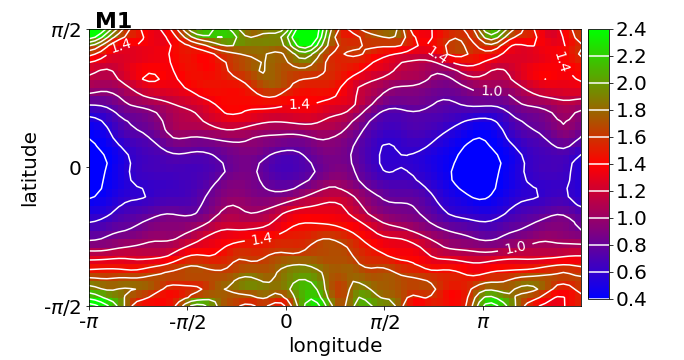} &
 \includegraphics[width=2.7in, trim=14 4 30 10, clip]{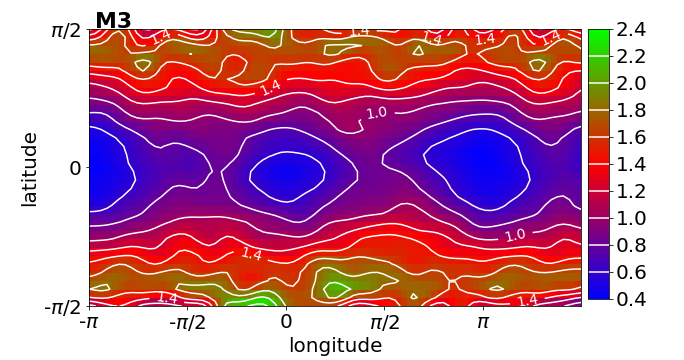} \\
  \includegraphics[width=2.7in, trim=14 4 30 10, clip]{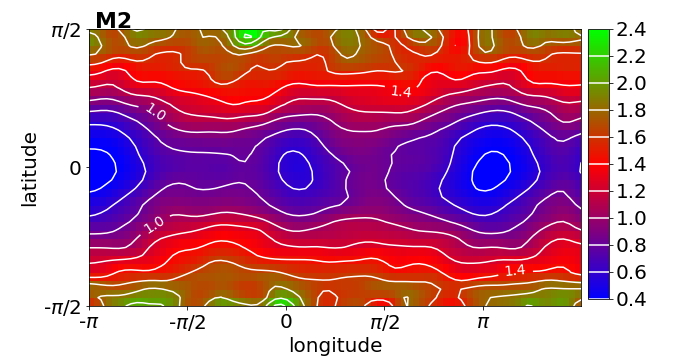} &
  \includegraphics[width=2.7in, trim=14 4 30 10, clip]{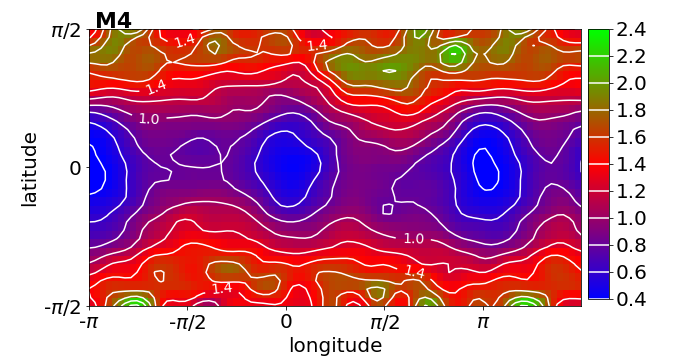} \\
  \includegraphics[width=2.7in, trim=14 4 30 10, clip]{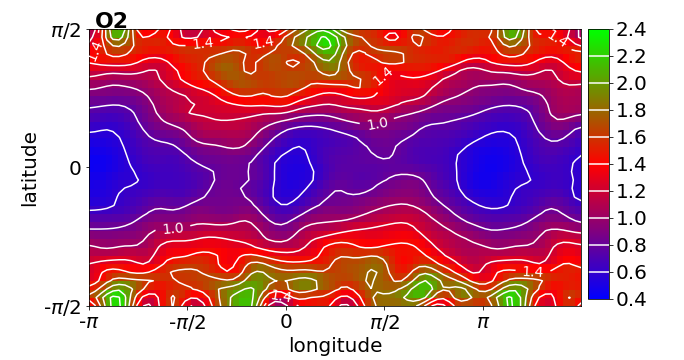} &
  \includegraphics[width=2.7in, trim=14 4 30 1, clip]{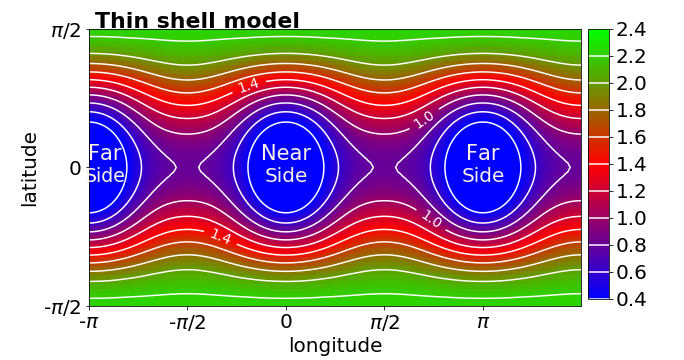} 
  \end{array}$$
  \vspace{-0.30in}
\caption{
Heat patterns in latitude (y-axis) vs longitude (x-axis) for spin synchronous bodies in  eccentric orbit. 
The point nearest to the perturber, the near side, is at longitude of 0, whereas longitude of $\pm \pi$
corresponds to the far side.  The longitude extends past $\pi$ on the right so that the far side can more easily
be seen.
We show the crustal tidal  heating rate per unit area integrated radially through the body.
Plotted is the tidal heating rate per unit area divided by its mean value (when integrated over the sphere). 
Simulations M1, M2, M3, M4, and O2 are shown on the top left, middle left, top right, middle right, and lower left, respectively, with  simulation names labelled on the top left of each panel.
On the lower right, we show the predicted normalized heating rate for eccentricity tides in a thin shell using 
the predicted function by \citet{beuthe13}. Near and far sides are labelled in white on this panel
The simulations have
the same spin and orbital period, but different tidal perturber masses and orbital semi-major axes. 
The M1--M4 simulations have spherical shells, whereas the shell in the O2 simulation is thinner on the poles
and on the  side nearest the perturber.
The orbital semi-major axes are 2, 4, 8,  16, and 4 Earth radii for the M1, M2, M3, M4, and O2 simulations, 
respectively (see Table \ref{tab:series}).
The polar heating pattern is typical of eccentricity tides.
Because of the proximity of the Earth,
the M1, M2 and O2 simulations (on the left) show asymmetric tidal heating with the nearer side (at a longitude of $0^\circ$
and to the left of  the center of each panel) heated
more than the far side (at a longitude of $-\pi$ or $\pi$ and on the sides of each panel).
  \label{fig:ll}}
 \end{figure*}
 
\begin{figure}
\centering
 \includegraphics[width=3in]{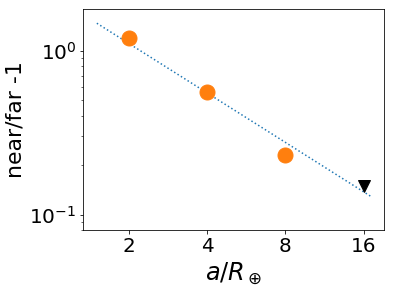} 
 \caption{The ratio of near to farside tidal heat flux subtracted by 1, for the M1--M4 simulations, 
 is plotted against the semi-major axis ratio.  The M4 simulation, shown as a black triangle, is an upper limit.
 The blue dotted line shows a linear dependence on the inverse of the semi-major axis.
 \label{fig:dots}
 }
\end{figure}

\subsection{Tidal heat distribution as a function of latitude and longitude}
 \label{sec:latlon}
 
Tidal heating rate  integrated from the heating rate per unit volume along radial rays 
and as a function of latitude and longitude are shown
in Figure \ref{fig:ll} for the 
same simulations we displayed in Figure \ref{fig:heat}. 
As heating may be asymmetric we show the entire range of longitude  $\phi \in [-\pi, \pi ]$.  
Longitude $0^\circ$ corresponds to the near side, facing the Earth, whereas $\phi = \pm \pi$ corresponds
to the far side.
Latitude is $ \pm \pi/2$ at the poles and zero at the equator.
Our $x$ axis extends past $\pi$ on the right so that the far side can be examined.
The region with $\phi > \pi $ on the right is the same as the region on the left with $\phi \in [ -\pi , -\pi/2]$.
Our plots differ from many studies that
 plot a restricted range with longitude  $\phi \in [-\pi/2, \pi/2]$ because of the expected  symmetry.  Usually 
 tidal heating from  the quadrupole potential term significantly dominates over the octupole term.
Heat maps in
Figure \ref{fig:ll} are normalized so that the average heating rate, computed from integrating over the surface,
is equal to 1. 

In Figure \ref{fig:ll} we  show on the lower right panel the heat pattern predicted with the thin shell model by \citet{beuthe13}.
We have computed the tidal heating rate per unit area for eccentricity tides using  associated Legendre polynomials, 
the $\Psi_C$ function, 
equations 34, 36, and the coefficients in Table 1 by \citet{beuthe13}.  
The heat pattern in our simulations with larger semi-major axis (the M3,M4 simulations) 
 resembles that  predicted for thin shells
(e.g., \citealt{peale78,ojakangas89,tobie05,beuthe13})
though the contrast between pole
and equator is somewhat larger than  in the predicted heat  map.  We attribute the difference to the coarseness of our simulation
(numbers of mass nodes) and to a free libration amplitude in the simulation that is not present in the predicted heat
distribution.   
The correspondence between predicted and numerically predicted heat maps demonstrates that the mass/spring model
can match tidally generated internal heat distributions.

Figure \ref{fig:ll} shows simulations that have the same spin and orbital periods.
However the perturber mass and semi-major axes differ in the M1--M4 simulations.   
For the more distant perturbers (M3, M4 simulations),
the tidal heating pattern is symmetric between near and far sides and resembles the heat flux
distribution predicted for a thin shell and eccentricity tides.  
However, the heating
pattern for the near and far sides differ for a closer perturber (the M1 and M2 simulations).   
The ratio of the heat flux on near and far sides is about $q_{t,{\rm near}}/q_{t, {\rm far}} \sim 1.5$
 for the M2 simulation with ratio of semi-major axis to body radius
 $a_o/R  = 15$, corresponding to $a_{\leftmoon}/R_\oplus = 4$  (using $R_\oplus/R_{\leftmoon} = 3.7$).
In section \ref{sec:tide} we estimated that an octupole tidal heating term would be smaller than
that of the dominant quadrupole by a factor of $R/a_o$ or 7\% for the M2 simulation.  
The contrast in the quadrupole heat distribution exceeds 2 so a somewhat larger contrast
than 7\% is not necessarily unexpected.

In Figure \ref{fig:dots} we plot the ratio of near to far side heat flux subtracted by 1,
$q_{t,{\rm near}}/q_{t, {\rm far}} - 1$, against orbital semi-major axis
for the M1--M4 simulations as a function of semi-major axis ratio, $a_{\leftmoon}/R_\oplus$.  
These are measured from averages of the heat maps in Figure \ref{fig:ll} in 
$10^\circ$ regions centered at latitude 0 and  longitudes $0$ and $ \pi$.
The blue dotted line in the plot is a linear
relation $q_{t,{\rm near}}/q_{t, {\rm far}} - 1 \propto a_o^{-1}$.  We plot an upper limit for the M4 simulation.
Comparison between M1--M4 simulations shows that the ratio of near and far side heat flux subtracted by 1
decreases approximately linearly 
as a function of inverse semi-major axis, or $\propto a_o^{-1}$. This is consistent with our expectation
that the asymmetry in heating arises from the octupole tidal term, as discussed in section \ref{sec:tide}.    
We attribute
the near/far side asymmetry in the tidal heating distributions in the M1 and M2
 simulations to the proximity of the Earth.

\subsection{Tidal heat distribution in a shell with uneven thickness}
 \label{sec:O2}

The lower left panel in Figure \ref{fig:ll} shows the O2 simulation that has an uneven thickness shell.  
Despite the extreme variations in shell thickness (see Figure \ref{fig:o2_lop}), 
the distribution of tidal heat flux integrated radially through the shell resembles that
of the other simulations.   This surprised us as we had expected the heating rates per unit volume to be similar.  
We show only a single lopsided shell simulation here, however we found that simulations with different shaped shell boundaries
and radial offsets and shell strengths 
exhibited similar behavior.  We find that the heat flux, or heat per unit area integrated through the shell, as a function 
of latitude and longitude,
is insensitive to shell thickness variations and is approximately proportional to the same function computed 
for a uniform thickness shell.

Recent analytical studies of elastic shells with variable thickness predict a phenomenon called {\it stress amplification} where
 stress is inversely proportional to the shell thickness \citep{behounkova17,beuthe18}.
Physically, thinner and weaker areas of the shell deform and dissipate more than thicker areas which are stiffer.
The insensitivity  of our simulated tidal heat distribution to shell thickness 
is consistent with this behavior.

For body with a crust or shell over a subsurface ocean, 
a coupling parameter, $R_{HS}$, is used to classify systems
as  being in a `hard-shell' or `soft-shell' limit (e.g., \citealt{goldreich10,beuthe18}).
Following equation 72 by \citet{beuthe18},
\begin{equation}
 R_{HS}\equiv \frac{E}{\rho g R} \frac{H}{R} = \frac{4 \pi}{3} \frac{E}{e_g} \frac{H}{R}
 \end{equation}
with $H$ mean shell thickness, $g = GM/R^2$ the surface gravitational acceleration and $e_g$ as defined in Equation \ref{eqn:eg}.
With $R_{HS} >1 $ a body is said to be in a hard-shell limit, with Enceladus having a hard shell due to its small size,
but Ganymede, Europa  \citep{A14} and the Moon when it had a magma ocean, in the soft-shell regime.
The parameter $e_g \propto \rho^2 R^2$  increases with radius, putting larger bodies with thin shells or crusts
in the soft-shell regime.  
The stress amplification phenomenon is predicted for hard shells,  such as Enceladus, and in
these the deformation of the surface depends on the depth integrated shear modulus
and the surface stresses  are inversely dependent on depth \citep{beuthe18}.

Our O2 simulation with thickness ratio $H/R = 0.3$ and $E_S/e_g = 7$ has $R_{HS} \sim 6 $ and is in the hard-shell regime.  
However during lunar magma
ocean solidification, the Moon, with $e_g \sim 40 $ GPa and  crustal thickness $H/R \sim $ 1 -- 2\%, has  $R_{HS} \sim 0.06$, and
would have been in the soft-shell regime.
We attempted to run simulations with thinner and softer shells, (so as to mimic the lunar crust). However, because the springs themselves
keep the simulated body from collapsing from self-gravity, we are limited to materials with Young's modulus $E/e_g>1$.  
To decrease the shell thickness we would
require more particles and shorter springs.   As the number of particles is proportional to the mean inter-particle
spacing to the third power, a significant reduction in shell thickness pushes us out of the realm of our code's current capabilities
(set primarily by the order $N^2$ direct gravity computation).
Springs not only connect shell nodes to shell nodes and core nodes to core nodes, but also connect shell
nodes to core nodes.   Our simulated shell base cannot slide on top of the core.  Even if we were able
to simulate thin shells, the simulated body  would not be consistent with hard shell floating on an ocean.

Previous computations of tidal heating in bodies that have a  shell over an internal ocean (such as Europa
or Enceladus) often assume a constant
 shell thickness  when computing the heating rate per unit volume 
(e.g., \citealt{peale78,ojakangas89,tobie05,wahr06,nimmo07,wahr09,beuthe13,beuthe15}).
This gives a pattern for the spatial heat distribution as a function of latitude and longitude
or a function $q_t(\vartheta,\phi)$ for the tidally generated heat per unit area as a function of latitude $\vartheta$
and longitude $\phi$.
Many studies necessarily have adopted the simple (but incorrect) assumption that
 $h$, the tidal heating rate per unit volume, follows the same spatial distribution in latitude and longitude as the $q_t$ that is 
  predicted from a constant thickness shell model.  Then viscoelastic tidal heat with a temperature dependent viscosity 
is assumed when computing the temperature profile through a conductive crust.
The result is a   crust or shell thickness as a function of latitude and longitude that is consistent with the depth
dependent tidal heating and the basal heat flux from the subsurface ocean
(e.g., \citealt{ojakangas89,tobie05,wahr06,nimmo07,wahr09}) but not necessarily with the elasticity of a variable thickness shell.

For soft shells, radial displacements due to tidal perturbation are set by the subsurface ocean
and are insensitive to shell thickness, however latitude and longitude dependent stress functions are still
dependent on shell thickness (see  \citealt{beuthe18}; section 5.2.4).
Is the soft shell regime consistent with tidal heating rate per unit volume  proportional to the tidal heating pattern 
predicted with a uniform thickness shell?    The shell must stretch and slide over an ocean surface that is 
a gravitational equipotential surface. 
We would guess that an integral of the total elastic energy in the shell would be minimized. 
Taking only the stress and strain components  tangential to the shell,  the total elastic energy  in the shell integrated
over its volume
 $U = \frac{1}{2} \int \epsilon_{ij} \sigma_{ij} \ dV \sim \frac{1}{2} \int \mu \epsilon_{ij} \epsilon_{ij} H\ d \Omega $, 
 where the integral on the right
 is over solid angle.
To minimize the total elastic energy from tidal deformation, stress and strain would tend to be lower where the shell is thickest. 
We lack simple analytical solutions and can't yet extend our simulations to cover the soft-shell regime, however
we suspect that in this regime too, the tidal heat distribution should depend on shell thickness and would be reduced
in thicker regions.

The insensitivity to thickness variations, exhibited by our simulations and predicted for the hard-shell regime, gives
a constraint on the tidally generated heat per unit volume;
\begin{equation}
 q_{t,{\rm uniform}}(\vartheta,\phi) \approx \int_{-H}^0 h_{\rm tidal} (\vartheta,\phi,z) dz 
 \end{equation}
where the function on the left is the heating pattern for of a uniform thickness shell and the
integral on the right is a function of depth in the shell.  The shell surface is at $z=0$
and the shell base is set by the shell thickness $H(\vartheta,\phi)$ as a function of latitude and longitude.
If the heating rate per unit volume $h$ is proportional to $q_t$ then a region with a thicker crust experiences more tidal heating.
The crust cannot continue to grow as its base then would start to melt.  However, if the heating rate per unit area is proportional to $q_t$
then thicker regions are cooler.  The crust in a cool region, such as on the far side of the Moon, 
can continue to grow.  

In summary, our simulations show that when the Moon was within a few Earth radii of the Earth, eccentricity tides
are asymmetric, with the far side experiencing less heating than the near side and poles.  
Simulations with perturbers at different orbital semi-major axes illustrate that the asymmetry is  dependent on the distance
between the Earth and the Moon.
The size of the asymmetry is approximately
consistent with that estimated from the ratio of the octupole gravitational component of the perturber to the quadrupole component.
A simulation with a variable shell thickness surprisingly showed a similar heating pattern to one with a uniform 
 thickness shell.   The insensitivity of the tidal heat flux to shell thickness implies that thicker areas of a shell
deform less than thinner regions, a phenomena dubbed  {\it stress amplification} by \citet{beuthe18}.
This phenomena is predicted for a hard-shell regime and our simulations also lie in this regime, however
the lunar crust during the epoch of magma ocean solidification is in a soft-shell regime.
We lack predictions for the sensitivity of heating distributions to thickness  (though see \citealt{beuthe18})
and the ability to simulate in the soft-shell regime, but we suspect that here too
crustal thickness variations would affect the tidal heating rate, with thicker regions less strongly tidally heated.
With both asymmetric heating and tidal heating rate per unit area insensitive to crustal thickness,
the lunar far side might form a thicker crust which could continue to grow. 
The asymmetry would persist as long as the Moon remains near the Earth and the octupole moment
is strong, and would rapidly diminish as the Moon recedes from the Earth. 

\begin{figure*}
\centering
    \includegraphics[width=6in]{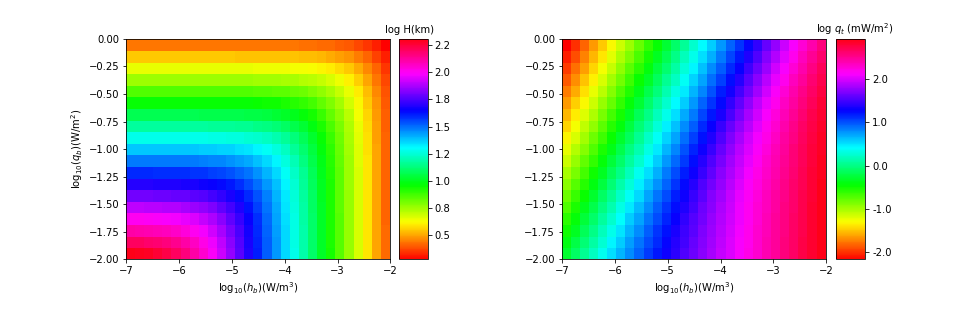} 
     \includegraphics[width=6in]{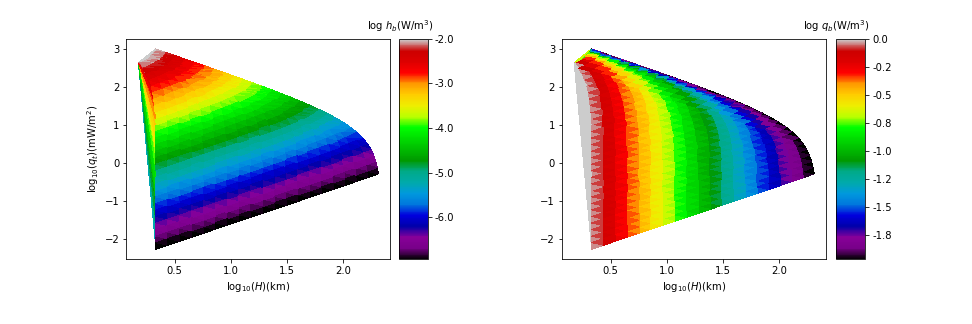} 
\caption{
Conductive thermal models.
 The top two panels have $x$ axis showing the  tidal heating rate per unit volume $h_b$  in W/m$^3$ and
 at the base of the lunar crust.  The $y$ axes shows the basal heat flux $q_b$ from the magma ocean into the crust
 in units of W/m$^2$. Both axes are logarithmic.   The top left panel colors give the crustal thickness $H$ 
 in km  and the right panel colors show the tidally generated heat integrated through the crust $q_t$
 in units of W/m$^2$.    The colorbar to the right of each plot shows numerical values for each color.
 These plots were made by numerically  integrating the heat diffusion equation (equation \ref{eqn:Tdiffusion}).
 The bottom two panels have $x$ axis the crustal thickness $H$ in km and $y$ axis the 
 tidal heat flux $q_t$ integrated through the crust in units of mW/m$^2$.  
 The bottom left panel shows the basal tidally generated heat per unit volume $h_b$ 
 and the bottom right panel shows the basal heat flux $q_b$. 
 The bottom two panels were constructed via triangulation from the points  in the top   two panels.  
The models used viscosity parameter $\gamma_Q (T_b - T_s) = 40$.  
The lower right panel shows that
at a crustal thickness $H$=10 km  and at a high tidal heat flux,  a small variation in the tidally generated
 heat flux $q_t$ gives large variations in the basal heat flux $q_b$.   
 If the basal heat flux at a particular latitude and longitude affects the crustal growth rates at the same
location, then variations in the  basal heat flux  could cause crustal growth rate variations. 
 \label{fig:lhq}}
\end{figure*}

\section{A tidally heated and conductive crustal lid}
\label{sec:asymm}

The asymmetry in the tidal heat flux between lunar near and far sides seen in our simulation 
ranges between 10-- 50\%  (expressed as a ratio in nearside to farside heat flux).
Can this weak asymmetry  contribute to uneven lunar crustal growth?  
A thermal model was used by \citet{garrickbethell10} to account
 for the thinner polar lunar crust compared to the far side equatorial value with a latitude dependent tidal heating distribution.  
We extend their model to take into account asymmetric tidal heating.
Parameters for our thermal models are listed in Table \ref{tab:moon} and additional symbols and
nomenclature are listed in Table \ref{tab:nomen}.
For the temperature at the lunar crust base, we take a magma ocean temperature of $T_b = 1175^\circ $ C, typical of
solidus (and following \citealt{garrickbethell10})
and a surface temperature of $T_s = 255^\circ$ K (typical of the lunar equator and following \citealt{vasavada99}).
  
We assume that the crust is in conductive thermal equilibrium with tidal heating in the crust and the
basal heat flux into the crust from the top of the lunar magma ocean.
We neglect radiogenic heating, as it is probably not an important contributor to the heat budget
during the early evolution of the Moon (see section 2 by \citealt{tian17}, or the supplemental
discussion by \citealt{garrickbethell10}).  
The time independent conductive heat diffusion equation for the crust
\begin{equation}
\frac{\partial }{\partial z} \left(K_T \frac{\partial T}{\partial z}  \right) + h_{\rm tidal}(T) = 0, \label{eqn:Tdiffusion}
\end{equation}
where $h_{\rm tidal}(T)$ is the tidal heating rate per unit volume.
The tidal heating rate should be a function of latitude, longitude and temperature.
Here  $K_T$ is the thermal conductivity which we assume is independent of temperature, $T$, or depth, $-z$.
Following \citet{garrickbethell10} we adopt a conductivity of $K_T = 1.8 {\rm W~m}^{-1} {\rm K}^{-1}$.
This value is an average of conductivity for feldspar rich plutonic rocks including anorthsite
over the temperature range of $200-750^\circ$ C \citep{clauser95}.

The basal  and surface heat fluxes
\begin{align}
q_b &\equiv  \left. -K_T \frac{\partial T}{\partial z}   \right|_{z=-H}  \label{eqn:qb}\\
q_s &\equiv  \left. -K_T \frac{\partial T}{\partial z}   \right|_{z=0}.  \label{eqn:qs}
\end{align}
The integrated tidal heat flux
\begin{equation}
q_t \equiv \int_{-H}^0 h_{\rm tidal}(T(z)) dz = q_s - q_b. \label{eqn:qt}
\end{equation}

Following previous studies \citep{ojakangas89,nimmo07,garrickbethell10},
rather than adopt an  Arrhenius relationship for the temperature dependence of the viscosity,  
we instead use the simplified Frank-Kamenetski approximation (e.g., \citealt{solomatov95});  
\begin{equation}
 \eta(T) = \eta_b \ e^{\gamma_Q (T_b-T)}  \label{eqn:visc}
\end{equation}
where $\eta_b$ is the reference viscosity at the crust base 
(at temperature $T_b$) and $\gamma_Q \equiv Q/R_gT_b^2$, where $Q$ is an activation energy and $R_g$ is the gas constant. 
The viscosity is high at low temperature and reaches a minimum of $\eta_b$ at the crustal base with temperature $T_b$ equal
to that of the well-mixed magma ocean.
Using $Q \sim 648 {\rm kJ~mol}^{-1}$ (a value for dry anorthsite; \citealt{rybacki00}) and gas
constant $R_g = 8.34 {\rm J~mol}^{-1} {\rm K}^{-1}$
(the nominal values from table S2 by \citealt{garrickbethell10}),
we compute  $\gamma_Q (T_b - T_s) = 44$.
With a Maxwell viscoelastic model and a viscosity estimated
for the lunar crust, the viscoelastic relaxation time $\tau_{\rm relax} = \eta/\mu$
is long compared to the tidal oscillation period.  In this limit, $\sigma_t \tau_{\rm relax} >1$, 
the tidal energy dissipation rate  per unit volume
$h_{\rm tidal} \propto \eta^{-1}$
and is maximum at the crust base.    With the viscosity following equation \ref{eqn:visc}, 
 the tidally generated heat per unit volume obeys 
\begin{equation}
h_{\rm tidal} (T) = h_b  \ e^{-\gamma_Q (T_b-T)},  \label{eqn:h_b}
\end{equation}
where $h_b$ is the maximum value at the crust base.

With an initial basal heat flux $q_b$ at temperature $T_b$ and basal tidal heating rate per unit volume $h_b$
we integrate the 1 dimensional conductive heat diffusion equation \ref{eqn:Tdiffusion} upward in $z$ until the temperature
reaches the surface temperature $T_s$.  We measure
the crustal thickness and integrated tidal flux using equation \ref{eqn:qt}.  We repeat this procedure for
a grid of $q_b$ and $h_b$ values.
We show in the top panels of Figure \ref{fig:lhq}
crustal thicknesses $H$ (top left, in km) and tidal flux $q_t$ (top right, in W/m$^2$) computed from these 1-dimensional  integrations.
The bottom two panels in Figure \ref{fig:lhq} were constructed via triangulation from the points shown in the top two panels.  
The points shown in the bottom two panels were then used to construct smooth interpolation functions giving $q_b(H,q_t)$ and $h_b(H,q_t)$.
We use these functions
 to estimate the basal heat flux $q_b$ and basal tidal heat per unit volume $h_b$ given 
a particular crustal thickness $H$ and integrated tidally heat flux $q_t$. 

The left hand side of the top left panel in Figure \ref{fig:lhq} shows that
at low levels of tidal heating, the crustal thickness depends inversely on the basal heat flux. This is expected
for a conductive crust with no additional heating as a high heat flux gives a high temperature gradient.    
At higher levels of tidal heating, the crustal thickness is reduced because
the tidal heat flux adds to the basal heat flux giving a higher heat flux through the crust.
The top right panel shows that higher level of basal tidal heating and higher level of tidal heating integrated through
the shell, $q_t$ (lower right on the top right panel)  
reduce the basal heat flux.  The tidal heating acts like a blanket, preventing the magma ocean from cooling.

An interesting region in the lower right panel of Figure \ref{fig:lhq} is at $H$ = 10 km and at
high tidal heat flux.  The basal heat flux is quite sensitive to the level of tidal heating.
Small variations in the tidal heating pattern could cause significant variations in the basal heat flux
and so on the rate of magma ocean crystallization below the crust.

\subsection{Estimates for crustal thickness variations}
\label{sec:grow}

\begin{figure*}
\centering
    \includegraphics[width=6in]{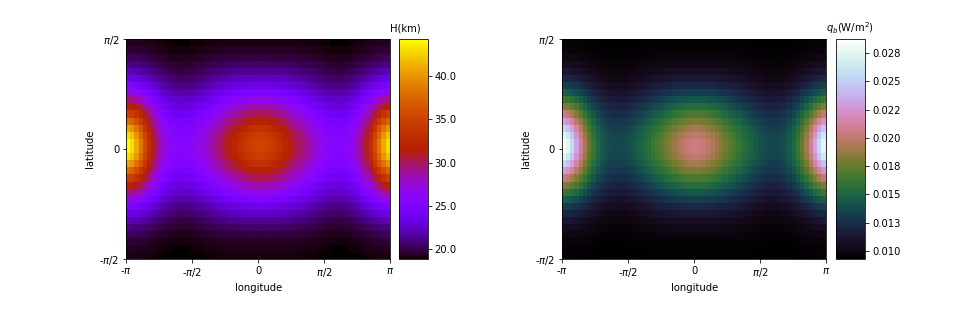} 
        \includegraphics[width=6in]{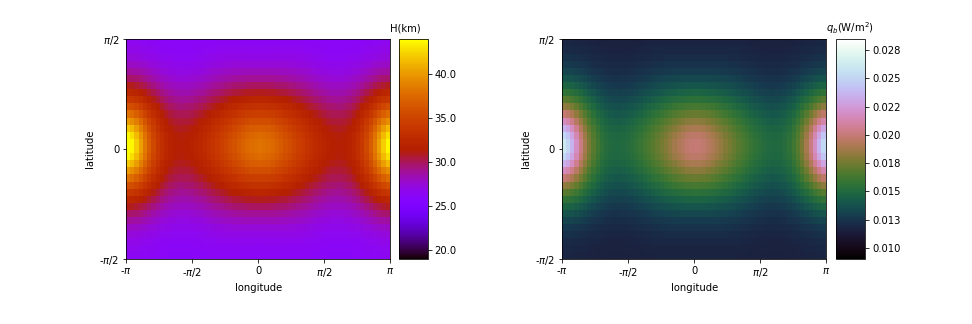} 
\caption{Two crustal growth models are used to estimate crustal thickness after $10^8$ years of growth.
In the top panels we use a crustal growth  model that assumes
an asymmetric tidal heat pattern for the tidal heat flux $q_t$ and crustal thickness as a function of latitude and longitude
 to estimate the crustal basal heat flux   $q_b$ as a function of latitude and longitude. 
In the bottom panels we assume that the tidal heat per unit volume $h_b$ depends on the asymmetric tidal heating pattern
 and use this and the crustal thickness as a function of latitude and longitude to estimate the basal heat flux, $q_b$.
The near side is at longitude of $0^\circ$ in the centre of each panel.
The crustal thickness distribution as a function of latitude and longitude is shown in panels on the left.
The  basal heat flux consistent with the crustal thickness in the left panels is shown on the right panels.
Color bars are the same for both growth  models.
In both growth models, the local basal heat flux is used to compute the growth rate of the crust.     
For both models, the crust is thicker on the far side than the near side, however the model with tidal heating setting
$q_t$ has enhanced asymmetry in the final crustal thickness.
 \label{fig:growcrust}}
\end{figure*}

We can estimate the growth rate of the crust with
\begin{equation}
\dot H =  \frac{f_{\rm plag}  }{\rho_{\rm plag} L_{\rm fusion}}.  \label{eqn:dotH}
\end{equation}
The factor $f_{\rm plag}$ is the proportion of plagioclase in the solidified portion of the melt.
We assume $f_{\rm plag} \sim 0.2$ \citep{snyder92,warren86} and adopt
 a latent heat of $L_{\rm fusion} = 4 \times 10^5$ J/kg for the magma ocean, following \citet{tian17}
 and a density of $\rho_{\rm plag} = 2670 {\rm km~m}^{-3}$.
 We only allow crustal growth, and do not allow previously grown crust to melt.

To estimate a crustal growth rate from a tidal heat distribution,  we make some additional
assumptions.    Starting with $H(\vartheta,\phi) = 10$ km,  we use either $q_t$ or $h_b$ to compute the
basal heat flux $q_b$.
We assume that the local basal heat flux sets the local crustal growth rate, ignoring mixing in the magma ocean.
The  crustal thickness is updated using equation \ref{eqn:dotH}.
The procedure is repeated, at each time step, updating the crustal thickness model.
Each growth model is integrated for $10^8$ years with final crustal thickness distribution and basal
heat flux at the end of the integration shown in the left and right panels of Figure \ref{fig:growcrust}.
The basal heat flux maps resembles the crustal thickness maps.  With higher basal heat flux beneath the
lunar far side, the magma ocean preferentially crystallizes below the moon's far side.

In our first growth model, shown in the top two panels of Figure \ref{fig:growcrust},
we assume that the distribution of tidal heating rate per unit area $q_t$ 
follows a pattern similar to those seen in our simulations. 
To match our asymmetric heating distribution we start with the  
heat distribution for a thin shell and eccentricity tide by \citet{beuthe13} and shown  in the lower right panel  of Figure \ref{fig:heat}.
To this we add a $ \cos 3\phi \cos \vartheta $ term to match the $\sim 20\%$ near and far side  asymmetry
we saw in our M2 simulation.
We assume that $q_t(\vartheta,\phi)$ follows this distribution, even though we allow the viscosity in the crust to depend on temperature.
We assume that the  average (over the sphere) tidal heating flux is $\bar q_t = 100 $ mW m$^{-2}$ and is constant in time.
From  $q_t(\vartheta,\phi)$ and $H(\vartheta,\phi)$ we use our interpolation function
$q_b(H,q_t)$ to compute the basal heat flux as a function of latitude and longitude.    

In our second growth model, 
 shown in the top two panels of Figure \ref{fig:growcrust},
 we assume that the tidal heating rate per unit volume at the crust base, 
$h_b$, follows the tidal heating pattern seen in our simulations.  This and the crustal thickness as a function
of latitude and longitude are used to compute the basal heat flux $q_b$, and the crust thickness increased using that value.
For this model we use a constant tidal heat per unit volume at the crust base of $h_b=10^{-4}{\rm  W~m}^{-3}$.

We found that the growth models were insensitive to the actual value used for $\gamma_Q$.
We have neglected surface temperature variations, though the poles could be $150^\circ$ cooler
than the equator \citep{vasavada99} and this could increase the thickness of the crust  
at the poles.

The model based on a tidal heat per unit area or $q_t$  distribution gives a 
 moderate crustal asymmetry with thickness $H \approx 32$ km on the near side and 44 km on the far side.
The thickness variations are larger for this model than for that based on the tidal heating rate per unit
volume at the crust base or $h_b$ distribution, with a similar far side thickness but a thicker near side at about 38 km.
We also explored a heating model with $h_b \propto \bar H/ H$ and with the same heat pattern, with 
a result similar to the constant $h_b$ model.

In these models we 
have maintained a constant rate of asymmetric tidal heating throughout crustal growth.   
Once the Earth's magma ocean solidifies,
 the semi-major axis of the Moon can more rapidly increase.   The orbital eccentricity can grow, so tidal heating
 could continue to be important (see \citealt{zahnle15}), however tidal heating would no longer be asymmetric.
 We explored removing the heating asymmetry at different times in the first growth model (based on $q_t$)
 finding that a moderate crustal thickness asymmetry persists when
 the  transition to quadrupolar heating takes place at greater than 40 Myr and is completely absent if
 the transition takes place earlier than 20 Myr.
 This time for crustal asymmetry to persist exceeds the time when  Earth's magma ocean solidifies, 
 which is at most 6 Myr \citep{zahnle15} and when the Moon would start to more rapidly drift away from the Earth.
Our simple crustal models predict that the asymmetry is wiped out once the tidal heating becomes entirely quadrupolar.

In summary, simple heat conductivity and crustal growth  models show 
that a moderate crustal asymmetry could persist during crustal growth, and if the thicker parts of the crust deform less
than the thinner parts, the asymmetry would be larger.
However, these simple heat conductivity models neglect mixing in the magma ocean and fail to match the extent of the observed lunar crustal thickness asymmetry.   Moreover, the crustal thickness asymmetry fails to persist once the 
heating becomes entirely quadrupolar.

\section{Summary and Discussion}
\label{sec:sum}

Recent work on the early evolution of the Earth \citep{zahnle07,sleep14,lupu14,zahnle15} have
shown that a magma ocean on the Earth may have reduced the tidally induced drift rate of the Moon's orbital semi-major
axis. This prolongs the Moon's
passage through the evection resonance and allows the lunar magma ocean to partially slowly solidify when
the Moon's orbit was eccentric and when it was quite near the Earth.  

We use a mass spring model viscoelastic code to model tidal heating of a spin synchronous Moon
in eccentric orbit about the Earth.  By tracking energy dissipation in the springs
we measure the distribution of internally dissipated heat. 
We ran simulations of a spinning body with a dissipative shell, mimicking an elastic lunar crust, overlaying a softer interior with lower viscoelastic
dissipation, mimicking a magma ocean.    
The distribution of tidally generated heat per unit area for perturbers at larger orbital semi-major axes
resembles that predicted for eccentricity tides with a thin hard shell model (that by \citealt{beuthe13}).

Our simulations show that when the Moon was within a few Earth radii of the Earth, eccentricity tides
are asymmetric.   The far side is less strongly heated than the near side and poles.  
The simulations show that the near/far side asymmetry 
decreases with increasing orbital  semi-major axis, and
the size of the asymmetry is 
consistent with that estimated from the ratio of the octupole gravitational component of the perturber to the quadrupole component.
The asymmetry is due to the octupole component of the gravitational potential of the perturber.  This term is only
significant when the perturber is nearby, and is usually neglected from tidal heating computations as a consequence.

A simulation with a variable shell thickness surprisingly showed a similar heating pattern to one with an
uniform thickness shell.   The insensitivity of the tidal heat flux to shell thickness could arise when thicker areas of a shell
deform less than thinner regions, a phenomena recently dubbed  {\it stress amplification} by \citet{beuthe18}.
Stress amplification is  predicted for bodies, such as Enceladus, that lie in a hard-shell regime \citep{beuthe18}.
During the epoch of lunar magma ocean solidification, the lunar crust would have been in a soft-shell regime.   
Unfortunately, we lack predictions for the sensitivity of heating distributions to thickness 
and the ability to simulate in the soft-shell regime, but we suspect that here too
crustal thickness variations would affect the tidal heating rate, with thicker regions less strongly heated.
Insensitivity of the heat flux distribution to thickness variations reduces the tidal heating rate in thicker crustal regions.
With both asymmetric heating and tidal heating rate per unit area insensitive to crustal thickness,
the lunar far side might form a stiff, thick and cool crust which could continue to grow.

We constructed  thermal conductivity models for the lunar crust to take into account 
the distribution of tidal heating. 
We used an asymmetric heating pattern like those seen in our  simulations as input to
the crustal growth model.   The local tidal heating rate affects the basal heat flux which
we used to compute the local crustal growth rate.    The basal heat flux sets the rate of magma ocean cooling and crystallization.
The resulting models illustrated moderate crustal thickness
variations after $10^8$ years of crustal growth, at a constant rate of tidal heating and with asymmetric tidal heating.   
Larger thickness variations were present at the end of crust growth for
a tidal heating rate per unit area independent of crustal thickness variations, than 
with heating rate per unit volume independent of crustal thickness variations.
However, thickness near/far side asymmetry did not persist in our simple models without asymmetric heating.
Thus once the moon moves away from the Earth, following solidification of the Earth's magma ocean,
the crustal thickness near/far side asymmetry is not maintained.   
Because recent scenarios exhibit a 
range of possible orbital eccentricity and semi-major axis evolutionary paths, our 
models also neglect variation in the tidal heating rate. Thus we have not taken into account  possible
of episodes of melting or that most crustal growth could have taken place 
when the Moon was at larger semi-major axis.

Our crustal growth models also neglect mixing in the magma ocean and fail to match the extent of the observed lunar
crustal thickness variations even if the tidal heating asymmetry is maintained.  
We assumed that the crust would grow where the basal heat flux
was high.    This makes the crust base akin to a refrigerated plate that is floating on ice-water, with 
ice more likely to collect on  the most strongly cooled regions of the plate. 
However, the magma ocean in the cooler regions could also be more turbulent and this could prevent accumulation of
floating plagioclase crystals (e.g., \citealt{tonks90}).   
Models of convection and crystal sedimentation in the lunar magma are  complex \citep{snyder92,solomatov00,parmentier02,nemchin09,lavorel09,ohtake12,laneuville13,charlier18}, 
but could in future explore sedimentation with uneven heat flux through the upper lunar magma ocean boundary.

The recent tidal heating models by \citet{garrickbethell14}, estimate the lunar tidal heat flux for an eccentricity 
of $e_{\leftmoon} =  0.02$ -- 0.03
and at a semi-major axis of $a_{\leftmoon}  = 20 R_\oplus$ (see their Table S13).  
Using the standard formula (e.g., equation 1 by \citealt{yoder81}) the  integrated tidal heating rate 
 is proportional to $e_{\leftmoon} ^2 a_{\leftmoon}^{-15/2}$.  
To achieve the same heating rate at a semi-major axis of  $8 R_\odot$ as at 20 $R_\oplus$, the eccentricity
must be 31 times lower, or  in the range $e_{\leftmoon} = $ 0.0006 -- 0.001.  For lunar crustal growth to take place when the
Moon was closer to the Earth,
the orbital eccentricity must be substantially lower than adopted by recent models. With the tethered Moon
scenario \citep{zahnle15}, the Moon does not acquire much eccentricity passing through the
evection resonance. These models have not yet estimated a range of possible values.   To estimate it, tidal evolution 
could be explored 
taking into account the degree of solidification and variations in the dissipation rates in both Earth and Moon.

Even if the early tidal heating asymmetry found here does not directly induce asymmetric crustal growth through variations
in basal heat flux, it might serve as a trigger for continued asymmetric crustal growth.  
If crustal growth is unstable (as discussed by \citealt{garrickbethell10}), 
with thicker and stronger regions tending to grow faster, as suggested by 
the stress amplification scenario for hard-shells \citep{behounkova17,beuthe18}, then early asymmetric tidal
heating might be amplified via later crustal growth,  after tidal heating becomes predominantly quadrupolar.
Alternatively, we could also conclude that an early episode of asymmetric tidal heating does not affect 
the later development of lunar crustal thickness variations.

We have explored  the possibility that the distribution of tidal heating could affect 
the difference in near and far side lunar crust thickness.  Our scenario is similar to
that proposed to explain thinner crust at the poles  by \citet{garrickbethell10}, (also see \citealt{garrickbethell14}) 
however our model
relies on proximity to the Earth as it depends on the strength of the Earth's octupole gravitational 
potential term.
Our mass spring simulations are not restricted to spherical symmetry and we used that
flexibility to search for asymmetric heating and sensitivity of the tidal heating pattern to crustal thickness variations.
The similarity between  predicted and simulated heat flux distributions suggests that 
this type of simulation, if expanded, may become increasingly flexible and powerful.
However, the simulations were difficult to adjust to minimize free libration,
and we often saw unintended polar wander and spin precession. 
Currently our simulations only simulate Kelvin-Voigt viscoelastic solids with a Poisson ratio of 1/4
and we cannot yet simulate other rheologies,  thinner crusts or hybrid models containing subsurface oceans.


\vskip 2 truein

Acknowledgements.

A.C. Quillen thanks L'Observatoire de la C\^ote d'Azur for their warm welcome
and hospitality March and April 2018.
This study was initiated with the help of Genn Shroeder.
We thank Patrick Michel, Mark Wieczorek, Marco Delbo, Randal Nelson,  
Mohamed Zaghoo, Cynthia Ebinger, Jing Luan,
Esteban Wright,
and Alexander Gysi for helpful discussions.

A. C. Quillen is grateful for generous support from the Simons Foundation.
This material is based upon work supported in part supported by NASA grant 80NSSC17K0771,
National Science Foundation Grant No. PHY-1757062 and  
NASA grant NNX15AI46G (PGGURP) to PI Tracy Gregg. 

Code used in this paper is available at \url{https://github.com/aquillen/moon_heat} .

\vskip 2 truein

{\bf Bibliography}

\bibliographystyle{elsarticle-harv}
\bibliography{moon_refs}

\end{document}